\documentclass{article}

\PassOptionsToPackage{numbers}{natbib}
\usepackage[main,final]{neurips_2026}

\usepackage[utf8]{inputenc} \usepackage[T1]{fontenc}    \usepackage{hyperref}       \hypersetup{hidelinks,pdfauthor={Mo Wang, Wenhao Ye, Zihan Ning, Jiayu Zuo, Junfeng Xia, Hongkai Wen, Quanying Liu}}
\usepackage{url}            \usepackage{booktabs}       \usepackage{amsfonts}       \usepackage{nicefrac}       \usepackage{microtype}      \usepackage[table]{xcolor}
\usepackage{courier}
\usepackage{amsmath}
\usepackage{graphicx}
\usepackage{wrapfig}
\usepackage{siunitx}
\usepackage{sidecap}
\usepackage{float}
\usepackage{algorithm}
\usepackage{subcaption}
\usepackage{multirow}
\usepackage{bm}
\usepackage{diagbox}

\definecolor{rowgray}{RGB}{242,242,242} 
\definecolor{ourred}{RGB}{199,106,122} 
\definecolor{ourblue}{RGB}{240,255,255} 
\newcommand{\heat}[2]{\cellcolor{ourred!#1}#2}

\newcommand{\best}[1]{\textbf{\textcolor{ourred}{#1}}}

\title{FlatClip: Reusing Image Foundation Models for fMRI Representation Learning via Cortical Flatmaps}

\author{  \textbf{Mo Wang}$^{1,2}$ \quad \textbf{Wenhao Ye}$^{1,3}$ \quad
  \textbf{Zihan Ning}$^{1}$ \quad \textbf{Jiayu Zuo}$^{1}$ \\
  \textbf{Junfeng Xia}$^{1}$ \quad \textbf{Hongkai Wen}$^{4,2,*}$ \quad
  \textbf{Quanying Liu}$^{1,*}$ \\[4pt]
  \normalfont\small $^{1}$Southern University of Science and Technology, Shenzhen, China \\
  \normalfont\small $^{2}$University of Warwick, Coventry, United Kingdom \\
  \normalfont\small $^{3}$Shenzhen University, Shenzhen, China \\
  \normalfont\small $^{4}$Fudan University, Shanghai, China \\[3pt]
  \normalfont\small $^{*}$Co-corresponding authors: \\
  \normalfont\small \texttt{hongkai.wen@gmail.com}, \texttt{liuqy@sustech.edu.cn}
}

\begin{document}

\maketitle

\begin{abstract}
Can the spatial organization of cortical activity help transfer frozen image representations to fMRI prediction? We introduce FlatClip, which renders cortical activity as anatomy-linked flatmaps and extracts features with a frozen SigLIP2 encoder. Only a lightweight downstream probe is trained, without fMRI-specific encoder pretraining or fine-tuning. Across four resting-state benchmarks, FlatClip outperforms the evaluated ROI-level baselines on HCP and ADNI, but is weaker on PPMI and below the strongest voxel-level models overall. On NSD visual-fMRI decoding, restricting inputs to visual or NSD-provided task-active cortex improves recognition over whole-cortex inputs. Spatial perturbations reduce performance with both retrained and fixed readouts, and anatomy-linked arrangements outperform cortical-pixel permutations across three colormaps. Together, these results demonstrate a practical route for reusing image-pretrained features in fMRI and highlight cortical arrangement and input-region selection as important design choices.
Code is available \href{https://github.com/OneMore1/FlatClip}{here}.
\end{abstract}

\section{Introduction}

\begin{figure}[t]
  \centering
  \includegraphics[width=0.55\textwidth]{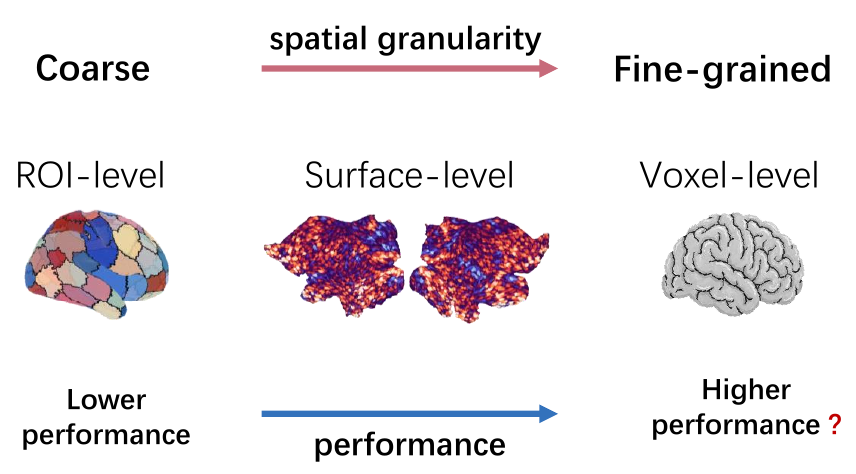}
\caption{
\textbf{Motivation.}
fMRI representations vary in spatial granularity.
FlatClip studies surface-level cortical flatmaps as a geometry-aware middle ground.
}
  \label{fig:motivation}
\end{figure}

Foundation models have recently become an important direction for fMRI representation learning. 
Instead of training a separate model for each dataset, task, or cognitive paradigm, fMRI foundation models aim to learn reusable brain representations from large-scale unlabeled neuroimaging data. 
Such models can reduce task-specific engineering and may improve transfer across heterogeneous downstream analyses~\cite{dong2024brain,caro2023brainlm,wang2026omni}. 
However, fMRI representation learning depends strongly on how brain activity is represented. 
Unlike natural images, fMRI is a high-dimensional 4D spatiotemporal signal with both spatial organization and temporal dynamics. 
Choosing the right spatial scale is therefore a central modeling decision (Fig.~\ref{fig:motivation}).

Most existing fMRI foundation models operate at either ROI level~\cite{dong2024brain,caro2023brainlm,yang2024brainmass,han2025hypergraph,wei2026large} or voxel level~\cite{wang2026omni,kim2023swift,wang2025slim,Wang2025TowardsAG,Sun2025VoxelLevelBS,Kwon2024PredictingTB}. 
ROI- and connectome-based models summarize voxel signals within atlas-defined regions and then model regional time series or inter-regional relationships. 
These representations are efficient and scalable, but they are coarse: they depend on parcellation choices and may discard fine-grained cortical geometry~\citep{wang2025dca,salehi2020there}. 
Voxel-based or volumetric models preserve more spatial detail and can capture local structure more directly, but they require specialized 3D/4D architectures, substantial computation, and storage of high-dimensional fMRI volumes. 
This trade-off raises a practical question: how effectively can a frozen image encoder use the spatial organization of cortical activity?
Pang et al. showed that macroscopic fMRI activity can be parsimoniously explained by geometric eigenmodes derived from brain shape, and that these modes explain spontaneous and task-evoked activity more effectively than connectome-derived modes~\cite{pang2023geometric}. 
This motivates representations that retain the spatial organization of cortical activity rather than reducing all signals to coarse regional averages.

We introduce \textbf{FlatClip}, a surface-level baseline that renders cortical activity as flatmap sequences, extracts features with a frozen SigLIP2 image encoder, and trains a lightweight downstream probe. The flatmaps provide a consistent anatomy-linked 2D+time layout between coarse ROI summaries and native volumes. Because the encoder requires no fMRI-specific pretraining or fine-tuning, FlatClip tests the predictive utility of cortical layouts using features learned outside neuroimaging.

We first evaluate FlatClip on four resting-state fMRI benchmarks, where subject-level prediction has no explicit stimulus region. 
Under a frozen-backbone, lightweight-probe protocol, FlatClip exceeds most ROI-level baselines while requiring no fMRI-specific encoder pretraining, although the strongest voxel-level models remain higher overall. 
We then evaluate visual-fMRI decoding on NSD COCO80, a stimulus-evoked setting in which the relevant cortical region is more explicit; NSD is a large-scale 7T fMRI dataset collected while subjects viewed natural images~\cite{allen2022massive}. 
Performance improves from whole-cortex input to HCP-MMP visual cortex and NSD-provided subject-level task-active regions, highlighting the importance of cortical input-region selection for visual decoding. 

Motivated by the resting-state and visual-fMRI observations, we conduct control experiments to test whether the gains are associated with spatial organization and spatial granularity. 
Geometry-disruption controls show that block-shuffling, cortical-pixel permutation, and FC heatmaps reduce performance relative to real flatmaps, indicating that the benefit is not merely due to image-like formatting. 
In a separate voxel-based comparison, adaptive patching improves HCP prediction, while its effect on AD diagnosis varies across metrics.
Together, these results support cortical spatial organization and task-relevant input selection as useful design choices for frozen-feature fMRI prediction.
Our main findings are summarized as follows:
\begin{itemize}
    \item We introduce FlatClip, a surface-level approach that reuses a frozen SigLIP2 image encoder for fMRI prediction through anatomy-linked cortical flatmaps, training only a lightweight downstream probe.

    \item Across four resting-state benchmarks, FlatClip outperforms the evaluated ROI-level baselines on HCP and ADNI, with task-dependent performance relative to voxel- and surface-based models.

    \item We show that cortical spatial arrangement and input-region selection affect the predictive utility of frozen image features, using spatial-perturbation controls, three colormaps, and comparisons across cortical masks.
\end{itemize}

\section{Related Work}

\paragraph{fMRI foundation models across representation scales.}
ROI- or atlas-based methods reduce voxel-level signals into region-wise time series or functional-connectivity graphs, and models such as BrainLM, Brain-JEPA, BrainMASS, LCM, and related graph-based approaches learn from ROI tokens, temporal dynamics, or graph-structured connectivity~\cite{dong2024brain,caro2023brainlm,yang2024brainmass,han2025hypergraph,wei2026large,xia2026brain}. 
At the other end of the scale, voxel- or volume-based models operate directly on 3D or 4D fMRI and preserve more fine-grained spatial structure~\cite{wang2026flexibrain,xia2026brain,kim2023swift,wang2026omni,wang2025slim,Wang2025TowardsAG,Sun2025VoxelLevelBS,Kwon2024PredictingTB,xia2026brainworld}. 
Our evaluations compare these pretrained representations with surface-level features under a frozen-backbone readout protocol.

\paragraph{Geometry-aware cortical surface and flatmap representations.}
Geometric and spectral analyses motivate retaining the spatial layout of cortical activity~\cite{pang2023geometric,behjat2020spectral}.
Pycortex is a standard toolbox for projecting volumetric neuroimaging data onto cortical surfaces and generating flattened cortical visualizations~\cite{gao2015pycortex}. 
Recent work has explored image-like or flatmap-based interfaces for fMRI, including visual decoding from fMRI signals and spatiotemporal masked-autoencoding on fMRI flatmap videos~\cite{Gao2024MinD3DAF,lane2025scaling}. 
FlatClip adopts this existing surface-rendering interface for frozen image-feature transfer.

\paragraph{Mesh- and sphere-based cortical models.}
Cortical learning also has a substantial literature beyond ROI and volumetric inputs. MoNet learns local filters on graphs and manifolds~\cite{monti2017monet}, while Spherical U-Net defines convolution and pooling on spherical cortical meshes~\cite{zhao2019spherical}. SiT represents cortical measurements as triangular spherical patches processed by a Transformer~\cite{dahan2022sit}. X-SiT and supervertex vision Transformers further investigate interpretable and multiscale surface representations for dementia analysis~\cite{bongratz2025xsit,baek2026supervertex}. CortexMAE-F instead pretrains a spatiotemporal encoder on fMRI flatmaps~\cite{lane2025scaling}. These approaches motivate our surface-model comparisons. FlatClip uses an existing image-pretrained encoder without fMRI-specific backbone training, offering a complementary reference for evaluating cortical representations.

\paragraph{Image-pretrained encoders as out-of-domain frozen baselines.}
Large-scale image-pretrained models have become reusable feature extractors for downstream vision tasks. 
Representative approaches include masked image modeling, contrastive image-text pretraining, self-distillation, and sigmoid-loss image--text pretraining, including MAE, CLIP, DINOv2, and SigLIP2~\cite{he2022masked,radford2021learning,oquab2023dinov2,tschannen2025siglip}. 
SigLIP2 extends the SigLIP family with captioning-based pretraining, self-supervised losses, and online data curation, improving visual representation transfer and dense-feature performance~\cite{tschannen2025siglip}. 

\section{Method}
\label{sec:method}

\subsection{Overview}

\begin{figure}[t]
\centering
\includegraphics[width=\linewidth]{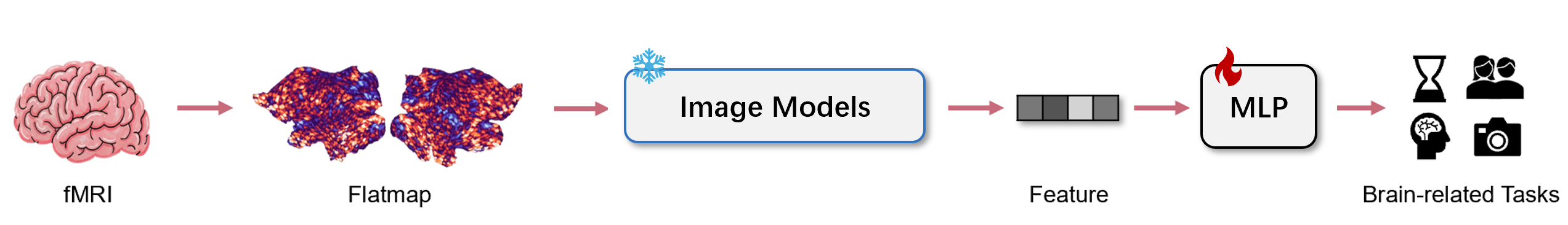}
\caption{\textbf{FlatClip pipeline.} Preprocessed cortical signals are rendered in a fixed surface layout and encoded by frozen SigLIP2. Resting-state prediction averages 40 frame features into one subject feature; NSD prediction uses a stimulus-level GLM response map. Only the downstream MLP is trained.}
\label{fig:flatclip_pipeline}
\end{figure}

Figure~\ref{fig:flatclip_pipeline} summarizes the three stages: cortical flatmap construction, frozen feature extraction, and downstream prediction. Resting-state inputs are sequences of time-point maps; NSD inputs are stimulus-level GLM response maps. The following sections specify their construction, pooling, and prediction losses.

\subsection{Geometry-aware Flatmap Sequence Adapter}
\label{sec:flatmap_adapter}

For each subject $i$, we construct a sequence of cortical flatmap frames
\[
\mathcal{F}_i = \{x_{i,1}, x_{i,2}, \ldots, x_{i,T}\},
\]
where each frame
\[
x_{i,t} \in \mathbb{R}^{H \times W \times C}
\]
is an RGB rendering ($C=3$) of the cortical activity at one time point. Starting from preprocessed volumetric fMRI, activity is projected onto cortical vertices; surface inputs already in fsLR space enter at this step. Pycortex rasterizes the vertex values onto a deterministic flatmap layout. We normalize and color-map the scalar values, apply the cortex or task-region mask, crop the rendered map, and composite the background onto RGB. Native cropped dimensions are reported in Appendix~\ref{app:added_dimensions}; the SigLIP2-NaFlex processor then prepares the encoder input.
By stacking frames along the temporal dimension, the sequence can be represented as a flatmap tensor
\[
\mathbf{X}_i = \mathrm{Stack}_{t=1}^{T}(x_{i,t})
\in \mathbb{R}^{T \times H \times W \times C}.
\]

For resting-state fMRI, $t$ indexes time points and the sequence length $T$ is fixed within each dataset. 
For visual-task fMRI, each sample corresponds to a stimulus-level response map estimated by a general linear model (GLM), where each response map is associated with one viewed image.

For NSD visual-fMRI decoding, we evaluate three cortical input regions: whole cortex, HCP-MMP visual cortex, and the NSD-provided subject-level task-active region, denoted as NSDgeneral. 
The HCP-MMP visual cortex mask is defined from visual ROIs in the HCP-MMP atlas in 32k fsLR space~\cite{glasser2016multi}. 
The NSDgeneral mask is provided by NSD and represents subject-level task-active cortex~\cite{allen2022massive}. 
These regions are rendered as flatmaps and processed using the SigLIP2-NaFlex input processor.

\subsection{Frozen SigLIP2 Encoder and Feature Construction}

Let $f_{\theta}$ denote the frozen SigLIP2 image encoder. 
For each flatmap frame, we extract one global image representation:
\[
z_{i,t}=f_{\theta}^{\mathrm{global}}(x_{i,t})\in\mathbb{R}^{d},
\]
where $f_{\theta}^{\mathrm{global}}(\cdot)$ denotes the global representation returned by the encoder. 
The encoder parameters $\theta$ are fixed throughout all experiments, and no gradient is propagated into the image encoder.

For resting-state fMRI, each subject has $T$ flatmap frames. 
We aggregate the frame-level global representations by average temporal pooling:
\[
h_i
=
\frac{1}{T}
\sum_{t=1}^{T} z_{i,t}
\in \mathbb{R}^{d}.
\]
In our main resting-state experiments, we use $T=40$ frames for each subject and $d=768$. 

For NSD, each sample is a stimulus-level GLM response map and uses patch-feature pooling rather than the encoder's global output. Let $P_{s,j}$ denote the valid patch features returned by the frozen vision encoder, arranged according to the processor's spatial shape. We remove trailing padding tokens and adaptively average-pool this feature grid to $16\times16$:
\[
Q_{s,j}=\operatorname{AdaptiveAvgPool}_{16\times16}(P_{s,j}),
\qquad
h_{s,j}=\frac{1}{256}\sum_{k=1}^{256}Q_{s,j,k}\in\mathbb{R}^{768}.
\]
The NaFlex processor allows up to 1,024 input patches in this extraction pipeline. The 256 pooled tokens are a post-encoder representation; their mean forms the main NSD feature, while token-rich variants retain all 256. Both tasks therefore provide a 768-dimensional main feature, with different pooling operations.

\subsection{Downstream Prediction}

For each downstream task, we train a lightweight probe $g_{\phi}$ on top of the frozen SigLIP2 feature. 
Only the probe parameters $\phi$ are optimized.

The main resting-state probe is an MLP with two hidden layers, $768\rightarrow256\rightarrow256\rightarrow C$, using BatchNorm, GELU, and dropout of 0.2. Features are standardized using training-set statistics. We use class-weighted cross-entropy, AdamW, a cosine learning-rate schedule, and seeds 42, 43, and 44; checkpoints are selected by validation weighted F1. Linear-probe sensitivity results are reported in Appendix~\ref{app:added_probes}. NSD uses a task-specific multilabel MLP and class-weighted binary cross-entropy.

For single-label resting-state classification, let
\[
\mathcal{D}_{\mathrm{rest}}=\{(h_i,y_i)\}_{i=1}^{N_{\mathrm{sub}}}
\]
denote the subject-level training set. 
The probe predicts
\[
\hat{y}_i=g_{\phi}(h_i),
\]
and is trained with class-weighted cross-entropy loss:
\[
\mathcal{L}_{\mathrm{CE}}
=
-\frac{1}{N_{\mathrm{sub}}}
\sum_{i=1}^{N_{\mathrm{sub}}}
\sum_{c=1}^{C}
\alpha_c\mathbf{1}[y_i=c]\log p_{\phi}(y_i=c\mid h_i),
\]
where $\alpha_c$ is the class weight estimated from training labels, $C$ is the number of classes, and $N_{\mathrm{sub}}$ is the number of training subjects.

For visual-fMRI recognition, let 
\(\mathcal{D}_{\mathrm{vis}}\) denote the set of subject--stimulus pairs \((s,j)\), and let 
\[
N_{\mathrm{vis}} = |\mathcal{D}_{\mathrm{vis}}|.
\]
Each pair has a feature \(h_{s,j}\) and a COCO80 multi-hot target
\[
y_j \in \{0,1\}^{C}, \quad C=80,
\]
where \(y_j\) depends only on the stimulus image and is shared across subjects. 
The lightweight probe outputs
\[
o_{s,j}=g_{\phi}(h_{s,j})\in\mathbb{R}^{C},
\]
and is trained with class-weighted binary cross-entropy:
\[
\mathcal{L}_{\mathrm{BCE}}
=
-\frac{1}{N_{\mathrm{vis}}}
\sum_{(s,j)\in\mathcal{D}_{\mathrm{vis}}}
\sum_{c=1}^{C}
\left[
w_c y_{j,c}\log \sigma(o_{s,j,c})
+
(1-y_{j,c})\log(1-\sigma(o_{s,j,c}))
\right],
\]
where \(\sigma(\cdot)\) is the sigmoid function and \(w_c\) is the positive-class weight computed from the training split.

\section{Evaluating Geometry-Aware Surface Representations}
\label{sec:forward_results}

\subsection{Resting-state fMRI Benchmarks}

We first evaluate FlatClip on four subject-level resting-state fMRI benchmarks: HCP sex classification, PPMI diagnosis prediction, ADNI MCI vs. CN classification, and ADNI AD vs. CN classification~\cite{jack2008alzheimer,marek2011parkinson,van2013wu}. 
These benchmarks cover demographic prediction in healthy adults and diagnostic-label prediction across neurodegenerative cohorts. 
All methods are evaluated on the same subject-level splits with frozen representations and lightweight downstream heads. Because the models have different native input requirements, the input construction follows each model's interface, while the readout protocol and splits are matched.
This provides a matched readout protocol for comparing frozen representations. 
Voxel-level models were evaluated using 40 input frames, following their model-specific requirements; FlatClip used the same temporal window for consistency. In contrast, ROI/FC-based models used approximately 200 frames for feature computation, according to the requirements of their respective pipelines.
We report Accuracy and weighted F1 against general-purpose fMRI foundation-model baselines, including ROI-level models (Brain-LM~\cite{caro2023brainlm}, BrainMASS~\cite{yang2024brainmass}, LCM~\cite{wei2026large}, Brain-Harmony~\cite{dong2025brain}) and voxel-level models (SwiFT~\cite{kim2023swift}, NeuroSTORM~\cite{Wang2025TowardsAG}, Omni-fMRI~\cite{wang2026omni}). 
Details about datasets, task design, ablation of time frame, and model-specific settings are provided in Appendix~\ref{app:task_details}.

As shown in Table~\ref{tab:rest_main}, FlatClip achieves competitive performance on HCP and ADNI without fMRI-specific encoder pretraining or end-to-end fine-tuning. 
It outperforms most ROI-level baselines while remaining below the strongest voxel-level models, consistent with its role as a surface-level middle-ground representation. 
This pattern establishes cortical flatmaps as a competitive surface-level representation under the evaluated frozen-backbone protocols. 
FlatClip is relatively weaker on PPMI, which may reflect a limitation of the current cortical-only input. 
Parkinson's disease involves subcortical and brainstem systems that are not fully captured by cortical surface flatmaps~\cite{braak2003staging,postuma2015mds}. 

\begin{table*}[t]
\centering 
\small
\setlength{\tabcolsep}{2pt} 
\renewcommand{\arraystretch}{1.2}
\caption{Performance on HCP, ADNI (MCI), ADNI (AD), and PPMI with Accuracy/weighted F1.
Darker cell color indicates stronger performance after column-wise min--max normalization.}
\label{tab:rest_main}

\resizebox{\textwidth}{!}{
\begin{tabular}{l cc cc cc cc}
\toprule 
\multirow{3}{*}{\diagbox[width=6em]{\textbf{Model}}{\textbf{Dataset}}} & 
\multicolumn{2}{c}{\textbf{HCP}} & 
\multicolumn{2}{c}{\textbf{ADNI (MCI)}} & 
\multicolumn{2}{c}{\textbf{ADNI (AD)}} &
\multicolumn{2}{c}{\textbf{PPMI}} \\

& \multicolumn{2}{c}{\textit{Sex Classif.}} & 
\multicolumn{2}{c}{\textit{Diagnosis}} & 
\multicolumn{2}{c}{\textit{Diagnosis}} &
\multicolumn{2}{c}{\textit{PD Diagnosis}} \\

\cmidrule(lr){2-3} \cmidrule(lr){4-5}\cmidrule(lr){6-7}\cmidrule(lr){8-9}
& \textbf{ACC $\uparrow$} & \textbf{wF1 $\uparrow$} 
& \textbf{ACC $\uparrow$} & \textbf{wF1 $\uparrow$} 
& \textbf{ACC $\uparrow$} & \textbf{wF1 $\uparrow$}
& \textbf{ACC $\uparrow$} & \textbf{wF1 $\uparrow$} \\
\midrule

Brain-LM
& \heat{0}{64.12$\pm$2.89} & \heat{0}{64.14$\pm$2.84}
& \heat{0}{45.86$\pm$5.18} & \heat{0}{45.78$\pm$5.12}
& \heat{22}{66.09$\pm$2.22} & \heat{23}{66.19$\pm$2.11}
& \heat{70}{62.50$\pm$4.89} & \heat{54}{57.15$\pm$6.62} \\

BrainMASS
& \heat{14}{68.24$\pm$1.29} & \heat{14}{68.18$\pm$1.36}
& \heat{33}{53.79$\pm$5.27} & \heat{34}{53.89$\pm$5.18}
& \heat{0}{60.87$\pm$5.32} & \heat{0}{60.91$\pm$5.44}
& \heat{15}{52.50$\pm$3.22} & \heat{24}{52.95$\pm$2.89} \\

LCM
& \heat{13}{67.94$\pm$3.23} & \heat{14}{67.93$\pm$3.26}
& \heat{30}{53.10$\pm$3.99} & \heat{22}{50.89$\pm$4.39}
& \heat{15}{64.35$\pm$3.53} & \heat{14}{64.05$\pm$3.82}
& \heat{50}{58.89$\pm$3.75} & \heat{0}{49.65$\pm$4.92} \\

Brain Harmony
& \heat{13}{67.84$\pm$2.56} & \heat{13}{67.73$\pm$2.57}
& \heat{52}{58.33$\pm$3.86} & \heat{53}{58.28$\pm$4.01}
& \heat{43}{70.90$\pm$5.28} & \heat{45}{71.04$\pm$5.35}
& \heat{59}{60.55$\pm$5.17} & \heat{56}{57.51$\pm$4.98} \\

\rowcolor{rowgray}
\textbf{FlatClip}
& \heat{62}{82.06$\pm$2.05} & \heat{64}{82.04$\pm$2.05}
& \heat{64}{61.11$\pm$1.48} & \heat{64}{60.75$\pm$1.16}
& \heat{62}{75.46$\pm$2.12} & \heat{64}{75.41$\pm$2.11}
& \heat{28}{54.86$\pm$1.96} & \heat{43}{55.72$\pm$1.63} \\

SwiFT
& \heat{76}{86.03$\pm$1.39} & \heat{74}{84.91$\pm$1.76}
& \heat{100}{69.78$\pm$4.63} & \heat{100}{69.31$\pm$5.23}
& \heat{59}{74.69$\pm$2.86} & \heat{61}{74.70$\pm$2.63}
& \heat{64}{61.55$\pm$2.52} & \heat{61}{58.23$\pm$4.20} \\

NeuroSTORM
& \heat{75}{85.72$\pm$1.44} & \heat{72}{84.36$\pm$2.00}
& \heat{48}{57.25$\pm$3.18} & \heat{47}{56.76$\pm$3.21}
& \heat{46}{71.64$\pm$0.66} & \heat{20}{65.35$\pm$2.06}
& \heat{93}{66.78$\pm$0.51} & \heat{55}{57.37$\pm$0.86} \\

Omni-fMRI
& \heat{100}{92.86$\pm$0.63} & \heat{100}{92.08$\pm$0.70}
& \heat{73}{63.40$\pm$2.55} & \heat{46}{56.67$\pm$9.60}
& \heat{100}{84.26$\pm$0.87} & \heat{100}{83.65$\pm$1.37}
& \heat{100}{68.13$\pm$0.60} & \heat{100}{63.65$\pm$2.04} \\

\bottomrule
\end{tabular}
}
\end{table*}

\begin{figure*}[t]
\centering
\includegraphics[width=\textwidth]{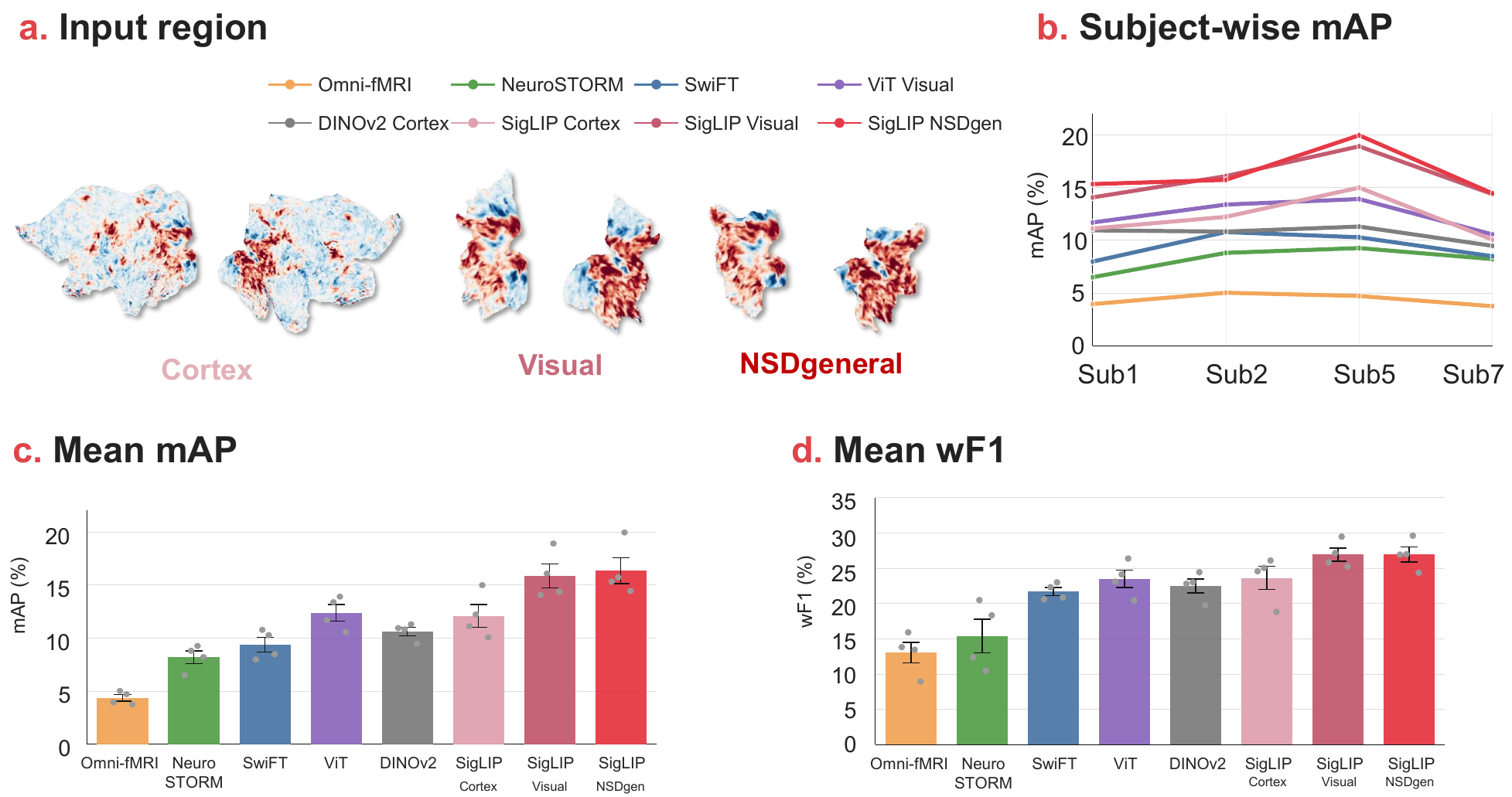}
\caption{Visual-fMRI COCO80 multi-label recognition on NSD.
(a) Illustration of the cortical input regions used by FlatClip: whole cortex, HCP-MMP visual cortex, and the NSD-provided subject-level task-active region (NSDgeneral).
(b) Subject-wise mAP across four NSD subjects for Omni-fMRI, NeuroSTORM, SwiFT and FlatClip variants.
(c,d) Mean mAP and weighted F1 across subjects.
All FlatClip variants use one global representation per GLM response map and the same downstream MLP classifier.
Dots denote individual subjects, and error bars denote SEM across subjects. 
Stimulus-level alignment statistics are reported in Appendix~\ref{sec:supp_image_fmri_alignment}.}
\label{fig:nsd_coco80_region}
\end{figure*}

\subsection{Visual-fMRI COCO80 Multi-label Recognition}
\label{sec:nsd_results}

We next evaluate FlatClip on stimulus-level visual-fMRI decoding using NSD COCO80 multi-label recognition. 
Each sample corresponds to a GLM-estimated cortical response map evoked by one NSD stimulus image, and the target is an 80-dimensional COCO multi-hot label vector. 
Unlike resting-state benchmarks, this setting is stimulus-evoked and has a more explicit task-relevant cortical substrate. 
It therefore allows us to test whether preserving cortical geometry is most useful when the retained geometry is relevant to the prediction target. 
We report within-subject decoding using mean average precision (mAP), averaged across the 80 COCO labels, and support-weighted F1, and compare FlatClip with general-purpose fMRI foundation-model baselines~\cite{kim2023swift,wang2026omni,Wang2025TowardsAG}.

For FlatClip variants, each GLM response map is encoded by the frozen SigLIP2 encoder into one global representation, followed by the same lightweight probe for COCO80 prediction. 
We vary the cortical input region across whole cortex, HCP-MMP visual cortex, and the NSD-provided subject-level task-active region, denoted NSDgeneral. 
The NSDgeneral setting is treated as a task-informed input-region setting. 
For voxel-level baselines on NSD, we follow each model's closest available visual-stimulus evaluation protocol. 
NeuroSTORM includes an NSD visual-stimulus evaluation, and Omni-fMRI provides an NSD-style protocol using repeated GLM response maps. 
For SwiFT, which supports variable-length inputs, each GLM response map is treated as a single-volume sequence. 

As shown in Fig.~\ref{fig:nsd_coco80_region}, FlatClip variants achieve stronger COCO80 recognition performance than the evaluated general-purpose fMRI foundation-model baselines. 
More importantly, performance improves as the input region becomes more specific to the visual task: whole-cortex input is weaker, HCP-MMP visual cortex performs better, and the NSDgeneral subject-level task-active region achieves the best mean mAP. 
For weighted F1, both visual-cortex and NSDgeneral inputs improve over whole-cortex input, while their difference is smaller. 
This pattern suggests that visual-fMRI decoding benefits not only from preserving cortical geometry, but also from selecting geometry that is relevant to the evoked visual task.
Additional image--fMRI feature-alignment analyses in Appendix~\ref{sec:supp_image_fmri_alignment} further show that SigLIP2-NSDgeneral features are more closely aligned with the corresponding image-feature geometry than both fMRI foundation-model baselines and broader FlatClip region choices. More details are provided in Appendix~\ref{sec:supp_ablation}.

\subsection{Geometry Controls and Representation Scale}
\label{sec:geometry_results}

We next test how cortical spatial arrangement affects the predictive utility of flatmap features.
We evaluate spatial perturbations of flatmap inputs and a separate comparison of ROI time-series and voxel-based models. 
Detailed control construction is provided in Appendix~\ref{app:perturbation}.

\paragraph{Geometry reduction.}
We perturb cortical flatmaps while keeping the SigLIP2 encoder frozen and the MLP architecture and training protocol fixed. An independent probe is trained for each input condition, with checkpoint selection on the corresponding validation set. These controls measure the task information recoverable from each perturbed representation.
Spatial block-shuffling permutes image blocks, while cortical-pixel permutation reassigns RGB values within the rendered cortical mask. Left-right hemisphere swap exchanges hemispheric image positions; within-hemisphere shuffling disrupts layout within each hemisphere. A further control randomizes spatial phase and rank-matches the foreground intensity histogram, with approximate frequency matching.
We also include FC heatmaps, random images, and a randomly initialized encoder as additional controls.

As shown in Table~\ref{tab:hcp_geometry_control}, real flatmaps achieve the best HCP sex-classification performance. 
The performance drops under spatial perturbations, especially within-hemisphere shuffling and phase-randomized, histogram-matched inputs, showing that anatomically arranged flatmaps yield more predictive features under the matched probe protocol.
FC heatmaps and random images perform much worse, and the randomly initialized encoder also degrades performance, supporting the utility of both spatially organized flatmap inputs and pretrained visual representations. 
DINOv2 controls show the same trend (Appendix~\ref{app:perturbation}).

\paragraph{ROI and voxel reference models.}
Table~\ref{tab:representation_comparison} compares a 450-ROI time-series model with native voxel models using standard or adaptive patching~\cite{wang2026omni}.
The voxel models obtain higher accuracy and weighted F1 than the ROI time-series model on both tasks.
Adaptive patching improves HCP accuracy and weighted F1; on AD diagnosis, accuracy increases slightly while weighted F1 decreases.
This comparison provides additional context for the evaluated representations and patching strategies, whose input formats and model architectures differ.

\begin{table}[t]
\centering
\small
\caption{\textbf{Geometry-control ablation on HCP sex classification.}
All input-control results use the frozen pretrained SigLIP2 encoder with one global representation per frame, except the explicitly labeled random-initialization control. The MLP architecture and training protocol are fixed, and a separate probe is trained for each condition.
Results are reported as mean$\pm$std over three seeds.
Significance markers indicate paired $t$-tests over the three probe-seed results, comparing Real flatmap against each control after Holm correction.
\best{Red} indicates the best performance.}
\label{tab:hcp_geometry_control}

\resizebox{\linewidth}{!}{
\begin{tabular}{l llll}
\toprule
\textbf{Control input} &
\textbf{ACC $\uparrow$} &
\textbf{wF1 $\uparrow$} &
\textbf{Balanced ACC $\uparrow$} &
\textbf{Macro F1 $\uparrow$} \\
\midrule
\rowcolor{rowgray}
Real flatmap
& \best{82.06\,${\pm}$\,2.05}
& \best{82.04\,${\pm}$\,2.05}
& \best{81.81\,${\pm}$\,2.03}
& \best{81.88\,${\pm}$\,2.06} \\

Spatial block-shuffle
& 76.14\,${\pm}$\,1.83$^{*}$
& 76.18\,${\pm}$\,1.83$^{*}$
& 76.21\,${\pm}$\,1.85$^{*}$
& 76.06\,${\pm}$\,1.84$^{*}$ \\

Left-right hemisphere swap
& 74.96\,${\pm}$\,1.17$^{*}$
& 74.89\,${\pm}$\,1.23$^{*}$
& 74.56\,${\pm}$\,1.31$^{*}$
& 74.64\,${\pm}$\,1.28$^{*}$\\

Within-hemisphere region shuffle
& 70.90\,${\pm}$\,1.63$^{*}$
& 70.87\,${\pm}$\,1.62$^{*}$
& 70.62\,${\pm}$\,1.61$^{*}$
& 70.64\,${\pm}$\,1.62$^{*}$ \\

Cortical-pixel permutation
& 72.93\,${\pm}$\,1.92$^{*}$
& 72.87\,${\pm}$\,1.97$^{*}$
& 72.58\,${\pm}$\,2.05$^{*}$
& 72.62\,${\pm}$\,2.02$^{*}$ \\

Real flatmap with random initialized
& 70.56\,${\pm}$\,1.34$^{*}$
& 70.56\,${\pm}$\,1.36$^{*}$ 
& 70.37\,${\pm}$\,1.41$^{*}$ 
& 70.35\,${\pm}$\,1.38$^{*}$ \\

Phase-randomized, histogram-matched
& 64.64\,${\pm}$\,2.29$^{*}$
& 64.49\,${\pm}$\,2.14$^{*}$
& 64.12\,${\pm}$\,2.03$^{*}$
& 64.14\,${\pm}$\,2.07$^{*}$ \\

FC heatmap
& 57.87\,${\pm}$\,2.69$^{*}$
& 57.93\,${\pm}$\,2.68$^{*}$
& 57.98\,${\pm}$\,2.52$^{*}$
& 57.80\,${\pm}$\,2.62$^{*}$ \\

Random image
& 48.90\,${\pm}$\,3.26$^{*}$
& 48.89\,${\pm}$\,3.42$^{*}$
& 48.69\,${\pm}$\,3.69$^{*}$
& 48.59\,${\pm}$\,3.65$^{*}$ \\
\bottomrule
\end{tabular}
}

\vspace{2pt}
\footnotesize{
Markers denote Holm-corrected paired tests against Real flatmap within each metric: $^{*}p<0.05$.
}
\end{table}

\begin{table*}[t]
\centering
\small
\caption{
\textbf{Comparison of ROI time-series and voxel-based models.}
The ROI input contains 450 regional time series; voxel models use native volumetric signals with standard or adaptive patching.
Results are Accuracy and weighted F1 (\%).
}
\label{tab:representation_comparison}

\resizebox{\textwidth}{!}{
\begin{tabular}{l l cc cc}
\toprule
\multirow{2}{*}{\textbf{Representation}} &
\multirow{2}{*}{\textbf{Resolution}} &
\multicolumn{2}{c}{\textbf{HCP Sex}} &
\multicolumn{2}{c}{\textbf{AD Diagnosis}} \\
\cmidrule(lr){3-4} \cmidrule(lr){5-6}
& &
\textbf{ACC $\uparrow$} & \textbf{wF1 $\uparrow$} &
\textbf{ACC $\uparrow$} & \textbf{wF1 $\uparrow$} \\
\midrule

ROI time series 
& 450 ROIs
& 65.68$\pm$4.68 & 65.51$\pm$4.47
& 60.65$\pm$5.61 & 60.75$\pm$3.90 \\

Voxel model
& Native 4D volume
& 80.74$\pm$2.41 & 80.32$\pm$3.21
& 72.22$\pm$10.2 & \best{71.66$\pm$9.54} \\

Voxel model
& Adaptive patches
& \best{82.81$\pm$0.57} & \best{82.88$\pm$0.55}
& \best{72.99$\pm$1.03} & 70.76$\pm$3.16 \\
\bottomrule
\end{tabular}
}
\end{table*}

\begin{table*}[t]
\centering
\small
\setlength{\tabcolsep}{2pt}
\renewcommand{\arraystretch}{1.2}
\caption{Implementation ablations for frozen image backbone, feature representation, and flatmap normalization. Results are reported as Accuracy / weighted F1. (\%) }
\label{tab:rest_token_ablation_siglip2}

\resizebox{\textwidth}{!}{
\begin{tabular}{l cc cc cc cc}
\toprule
\multirow{3}{*}{\diagbox[width=8em]{\textbf{Setting}}{\textbf{Dataset}}} &
\multicolumn{2}{c}{\textbf{HCP}} &
\multicolumn{2}{c}{\textbf{ADNI (MCI)}} &
\multicolumn{2}{c}{\textbf{ADNI (AD)}} &
\multicolumn{2}{c}{\textbf{PPMI}} \\

& \multicolumn{2}{c}{\textit{Sex Classif.}} &
\multicolumn{2}{c}{\textit{Diagnosis}} &
\multicolumn{2}{c}{\textit{Diagnosis}} &
\multicolumn{2}{c}{\textit{PD Diagnosis}} \\

\cmidrule(lr){2-3} \cmidrule(lr){4-5}\cmidrule(lr){6-7}\cmidrule(lr){8-9}
& \textbf{ACC $\uparrow$} & \textbf{wF1 $\uparrow$}
& \textbf{ACC $\uparrow$} & \textbf{wF1 $\uparrow$}
& \textbf{ACC $\uparrow$} & \textbf{wF1 $\uparrow$}
& \textbf{ACC $\uparrow$} & \textbf{wF1 $\uparrow$} \\
\midrule

DINOv2
& 80.37$\pm$2.55 & 80.33$\pm$2.53
& 60.68$\pm$2.67 & 60.44$\pm$1.98
& 75.46$\pm$4.01 & 75.20$\pm$3.69
& 51.39$\pm$3.18 & 51.58$\pm$2.25 \\

ViT-B
& 80.10$\pm$1.71 & 80.12$\pm$1.68
& 54.48$\pm$2.80 & 54.82$\pm$3.18
& 76.23$\pm$4.06 & \best{76.05$\pm$3.88}
& 56.67$\pm$4.51 & 54.03$\pm$5.19 \\

\rowcolor{rowgray}
SigLIP2
& \best{82.06$\pm$2.05} & \best{82.04$\pm$2.05}
& \best{61.11$\pm$1.48} & \best{60.75$\pm$1.16}
& \best{75.46$\pm$2.12} & 75.41$\pm$2.11
& \best{54.86$\pm$1.96} & \best{55.72$\pm$1.63} \\

\midrule
40 Patch-Mean
& \best{84.94$\pm$1.06} & \best{84.91$\pm$1.06}
& 58.12$\pm$1.96 & 57.82$\pm$1.87
& \best{76.39$\pm$4.17} & \best{76.36$\pm$3.93}
& \best{56.94$\pm$4.21} & \best{56.48$\pm$4.27} \\

40 Global tokens
& 83.08$\pm$0.59 & 83.02$\pm$0.66
& 60.68$\pm$1.96 & 60.02$\pm$2.07
& 74.07$\pm$4.88 & 74.13$\pm$4.98
& 53.47$\pm$7.54 & 53.93$\pm$6.14 \\

\rowcolor{rowgray}
One Global token
& 82.06$\pm$2.05 & 82.04$\pm$2.05
& \best{61.11$\pm$1.48} & \best{60.75$\pm$1.16}
& 75.46$\pm$2.12 & 75.41$\pm$2.11
& 54.86$\pm$1.96 & 55.72$\pm$1.63 \\

\midrule

Frame Z-Score
& \best{82.23$\pm$1.34} & \best{82.23$\pm$1.28}
& 58.12$\pm$3.23 & 58.28$\pm$3.27
& 73.15$\pm$0.80 & 73.52$\pm$0.91
& 49.65$\pm$3.35 & 50.68$\pm$3.10 \\

Voxel Z-Score
& 74.49$\pm$2.20 & 74.52$\pm$2.22
& 58.55$\pm$0.60 & 56.22$\pm$1.28
& \best{80.56$\pm$1.96} & \best{79.68$\pm$2.33}
& 50.35$\pm$2.60 & 51.17$\pm$0.76 \\

\rowcolor{rowgray}
Global Z-Score
& 82.06$\pm$2.05 & 82.04$\pm$2.05
& \best{61.11$\pm$1.48} & \best{60.75$\pm$1.16}
& 75.46$\pm$2.12 & 75.41$\pm$2.11
& \best{54.86$\pm$1.96} & \best{55.72$\pm$1.63} \\

\bottomrule
\end{tabular}
}
\end{table*}

\subsection{Implementation Ablations}
\label{sec:ablation_results}

Table~\ref{tab:rest_token_ablation_siglip2} evaluates the image backbone, feature representation, and flatmap-value normalization. Backbone comparisons fix the input and readout; SigLIP2 is used as the default encoder. Retaining additional frame or patch information improves some tasks, while the main setting uses one compact subject-level feature.
Detailed backbone and visual-fMRI ablations are provided in Appendices~\ref{app:task_details} and~\ref{sec:supp_ablation}.

\subsection{Surface-Model Comparisons}

Table~\ref{tab:added_surface_models} compares FlatClip with frozen CortexMAE-F and an end-to-end SiT-style parcel-token surface Transformer under the corresponding downstream splits. CortexMAE-F uses the released frozen checkpoint with a short-window adaptation of the first 40 source frames. CortexMAE-F is slightly stronger on HCP, while FlatClip achieves higher accuracy and weighted F1 on both ADNI tasks. The SiT-style model is competitive on ADNI AD/CN. Appendix~\ref{app:added_surface} specifies the checkpoints and surface-input construction.

\begin{table}[t]
\centering\small
\caption{\textbf{Comparison with surface models.} Accuracy / weighted F1 (\%), mean$\pm$std over three runs. CortexMAE-F pretraining includes HCP. The SiT-style column reports the parcel-token configuration; implementation details are given in Appendix~\ref{app:added_surface}.}
\label{tab:added_surface_models}
\resizebox{\linewidth}{!}{\begin{tabular}{lccc}
\toprule
Task & FlatClip & CortexMAE-F & SiT-style parcel-token surface \\
\midrule
HCP sex & $82.06\pm2.05$ / $82.04\pm2.05$ & $\mathbf{82.40\pm1.27}$ / $\mathbf{82.37\pm1.26}$ & $55.84\pm5.19$ / $55.74\pm5.15$ \\
ADNI AD/CN & $\mathbf{75.46\pm2.12}$ / $\mathbf{75.41\pm2.11}$ & $69.91\pm6.26$ / $70.43\pm6.31$ & $74.54\pm2.85$ / $73.07\pm2.90$ \\
ADNI MCI/CN & $\mathbf{61.11\pm1.48}$ / $\mathbf{60.75\pm1.16}$ & $58.12\pm0.74$ / $57.46\pm1.46$ & $55.98\pm3.96$ / $55.59\pm2.79$ \\
\bottomrule
\end{tabular}}
\end{table}

\subsection{Spatial Arrangement Across Renderers and Readouts}

We extend the spatial-perturbation experiments with a $2\times3$ geometry-by-rendering design. All conditions use a fixed $224\times224$ canvas, the same HCP split and 40-frame inputs, frozen SigLIP2, and the same MLP protocol. Each renderer is evaluated with real flatmaps and the same fixed cortical-pixel permutation, with a freshly trained probe for every condition. Table~\ref{tab:added_renderers} shows a consistent weighted-F1 advantage for the real arrangement across all three colormaps.

Fixed readouts trained on real flatmaps degrade substantially under block shuffle, cortical-pixel permutation, and phase randomization. Retraining the linear probe recovers part of the performance but remains below the real-flatmap result. Appendix~\ref{app:added_readout} details these complementary tests of recoverable information and decision-boundary transfer.

\begin{table}[t]
\centering\small
\caption{\textbf{Geometry effect across renderers.} Weighted-F1 gain (percentage points) of real flatmaps over the corresponding fixed cortical-pixel permutation.}
\label{tab:added_renderers}
\begin{tabular}{lc}
\toprule
Renderer & Real minus permuted wF1 \\
\midrule
Inferno & $+7.26$ \\
Grayscale & $+8.24$ \\
RdBu\_r & $+8.74$ \\
\bottomrule
\end{tabular}
\end{table}

\subsection{Measured Covariates, Temporal Sampling, and Probe Choice}

We assess the sensitivity of frozen FlatClip features to available demographic, motion, site, acquisition, and cohort variables. Covariate models are fitted on training subjects and applied unchanged to validation and test subjects before probe training. HCP weighted F1 is 80.88 after age adjustment, compared with 82.04 for the raw features. On ADNI AD/CN, motion adjustment yields 75.34 weighted F1, compared with 75.41 before adjustment. Joint adjustments produce task-dependent changes, reported in Appendix~\ref{app:added_confounds}.

Alternative non-overlapping 40-TR windows retain predictive performance on the ADNI tasks: MCI/CN weighted F1 is 62.69 and 63.20 in the paired window evaluation, and AD/CN weighted F1 is 75.41 and 73.35 (Appendix~\ref{app:added_windows}). Linear probing reaches 84.77 accuracy and weighted F1 on HCP, while the MLP is stronger on the ADNI and PPMI tasks. Increasing the number of labeled HCP training subjects improves both readouts, with MLP weighted F1 rising from 69.92 at 10\% of the training set to 82.04 at 100\% (Appendix~\ref{app:added_scaling}).

\subsection{Direct-Volume Rendering Comparisons}

We additionally apply frozen SigLIP2 to 2D renderings derived directly from acquired volumes. On HCP, axial mean projection gives 72.81 weighted F1, whereas independently encoding the non-empty slices of each volume and averaging their embeddings across slices and time gives 89.32. FlatClip gives 82.04 with one rendered cortical image per time point. These results illustrate the performance and image-count trade-off between the input representations. On NSD subj01, direct-volume slice encoding gives 4.342 mAP and 15.922 weighted F1, compared with the reported FlatClip-NSDgeneral result of 15.311 and 26.922. Full rendering and pooling details are provided in Appendix~\ref{app:added_volume}.

\paragraph{Native temporal inputs to Omni-fMRI.}
We also evaluate the released Omni-fMRI encoder on NSD subj01 raw BOLD using continuous 40-TR windows and event-centered stacks. These inputs yield 3.804/16.531 and 3.840/16.188 mAP/weighted F1, respectively. The two temporal constructions perform similarly, with continuous windows giving slightly higher weighted F1. Appendix~\ref{app:added_omni} provides the segment construction, data split, and full results.

\section{Discussion}

The experiments highlight two useful design choices: retaining an anatomy-linked spatial arrangement and selecting cortical inputs relevant to the task. Spatial perturbations assess sensitivity to cortical arrangement, while comparisons across NSD input masks examine input-region selection. Same-mask baselines compare representations and readouts under matched cortical coverage. Together, these analyses show how spatial arrangement and cortical coverage influence prediction with frozen image features.

FlatClip also separates reusable feature extraction from task-specific readout. Different cortical inputs can be evaluated with the same frozen encoder. This makes FlatClip a practical reference without fMRI-specific backbone training for comparisons with domain-specific models. More broadly, the results highlight cortical input design as a means of reusing pretrained visual representations, complementing efforts to develop fMRI-specific encoders.

\paragraph{Limitations.}
Flatmaps introduce cuts and metric distortion and omit subcortical and brainstem signals. Temporal mean pooling also discards frame order. Future work will examine hybrid cortical--subcortical inputs and explicit temporal modeling, including naturalistic movie paradigms.

\newpage
\appendix

\section{Details of Rest fMRI experiments}
\label{app:task_details}

\subsection{Details of Datasets and Task}

\paragraph{Human Connectome Project (HCP).}
The HCP FlatClip feature cohort contains 1,000 subjects with matched features and labels, partitioned into 602 training, 201 validation, and 197 test subjects. Table~\ref{tab:added_classes} gives the class counts for this partition, which is also used by the supplementary HCP feature evaluations.

\paragraph{Parkinson’s Progression Markers Initiative (PPMI).}
PPMI is a longitudinal observational study of Parkinson's disease progression~\cite{marek2011parkinson}. Our three-way classification cohort contains 474 participants: 29 controls, 142 with PD, and 303 prodromal participants. The training, validation, and test partitions contain 331, 47, and 96 participants, respectively.

\paragraph{Alzheimer's Disease Neuroimaging Initiative (ADNI).}
The Alzheimer's Disease Neuroimaging Initiative (ADNI) is a longitudinal study of Alzheimer's disease progression~\cite{jack2008alzheimer}. The MCI/CN task uses 381 subjects, split into 266/37/78 training/validation/test subjects; AD/CN uses 350 subjects, split into 244/34/72. The alternative-window analysis is described in Appendix~\ref{app:added_windows}.

\subsection{Evaluation Protocol}

Evaluation uses subject-level partitions, with all frames from a subject assigned to one partition. Models use the frozen-feature protocol in Section~\ref{sec:method}; checkpoints are selected on validation data and evaluated on the held-out test set. Supplementary probe experiments keep these partitions fixed and report mean and standard deviation over seeds 42, 43, and 44.

\subsection{Details of Baseline Models}
\label{app:baseline}
In this section, we introduce our baseline models.
 
\paragraph{BrainLM} is the first fMRI foundation model specifically designed to capture the spatiotemporal dynamics of brain activity through a Transformer-based masked autoencoder architecture. It employs the AAL-424 atlas for the parcellation of brain regions, transforming fMRI recordings into a 424-dimensional set of ROIs. The model is trained on an extensive dataset consisting of 6,700 hours of fMRI data derived from 77,298 recordings across the UK Biobank and the Human Connectome Project. 
 
\paragraph{BrainMASS} uses Schaefer 100-region brain networks and combines masked ROI modeling with latent representation alignment. The study collection comprises 70,781 samples from 46,686 participants across 30 datasets, of which 64,584 samples are used for pretraining~\cite{yang2024brainmass}.

\paragraph{Brain-Harmony} is a multimodal brain foundation model designed to jointly represent brain morphology and function within a unified one-dimensional token space. The model first encodes functional MRI ROI time-series and structural T1-weighted MRI into modality-specific token embeddings, and then uses a Transformer-based harmonizer to fuse these tokens into a shared subject-level representation. 
In our baseline setting, we use the official pretrained Brain-Harmony backbone with a ViT-Base architecture and 128 latent tokens. 

\paragraph{LCM} is an fMRI foundation model for connectome-based brain representation learning, which formulates pretraining as a multitask learning problem by leveraging brain--environment interaction variables, including demographic and behavioral targets, together with large-scale functional neuroimaging data. 
Its architecture represents brain connectomes as node-level embeddings and uses a decoder-style prediction head with learnable task queries to support multiple downstream objectives such as sex prediction, behavioral recognition, and disease classification. 

\paragraph{SwiFT (Swin 4D fMRI Transformer)} is an end-to-end framework for modeling high-dimensional spatiotemporal fMRI dynamics directly from raw 4D volumes, avoiding the information loss that can arise from hand-crafted features or atlas-level summaries. 
It adopts a computationally efficient 4D shifted-window attention design to capture local and long-range spatiotemporal dependencies in brain activity. 
SwiFT also supports contrastive self-supervised pretraining, which further improves downstream transfer performance on large-scale fMRI benchmarks. 

\paragraph{NeuroSTORM} is a general-purpose neuroimaging foundation model designed to learn representations directly from raw 4D fMRI volumes. 
Its backbone is based on Shifted-Window Mamba (SWM), which is used to model long-range spatiotemporal dependencies while maintaining computational efficiency. 
NeuroSTORM is pre-trained on large-scale fMRI data from more than 50,000 subjects across UK Biobank, ABCD, and HCP, and introduces a Spatiotemporal Redundancy Dropout (STRD) module to reduce redundant temporal information during representation learning. 
This design enables NeuroSTORM to serve as a transferable foundation model for downstream fMRI prediction tasks.

\paragraph{Omni-fMRI} is an atlas-free fMRI foundation model designed to learn directly from voxel-level signals, avoiding the information loss and atlas-dependent biases introduced by predefined region-level parcellations. 
To scale pretraining to 49,497 fMRI sessions from nine datasets, Omni-fMRI introduces a dynamic patching mechanism that reduces computational cost while preserving informative spatial structure. 
The model is evaluated through a comprehensive benchmark suite covering 11 datasets and diverse resting-state and task-based fMRI tasks, where it consistently outperforms existing fMRI foundation models. 
In our experiments, we use Omni-fMRI as a strong voxel-level foundation-model baseline for assessing whether FlatClip can achieve competitive transfer performance without fMRI-specific encoder pretraining.

\subsection{Ablation of Architectures}

We further compare different frozen image backbones and feature extraction choices for resting-state prediction (Table~\ref{tab:rest_backbone_token_ablation}). 
For the main comparisons, all methods use one global  representation and the same downstream MLP probe to ensure a fair evaluation of the frozen backbone. 
This setting follows the standard use of global or \texttt{[CLS]} representations in ViT-style encoders for image-level prediction, where the global token summarizes the input image into a single feature vector.  The ViT baseline uses a standard ImageNet-pretrained ViT-B/16 visual encoder from torchvision. 
Across backbones, SigLIP2 achieves the strongest performance on HCP  and is used as the default backbone in subsequent geometry-control analyses. 
The frame-token and patch-mean variants assess the effect of retaining additional feature information.

\begin{table*}[t]
\centering
\small
\setlength{\tabcolsep}{2pt}
\renewcommand{\arraystretch}{1.2}
\caption{Ablation of frozen image backbone and feature extraction strategy. 
Results are reported as Accuracy / weighted F1.
}
\label{tab:rest_backbone_token_ablation}

\resizebox{\textwidth}{!}{
\begin{tabular}{l l cc cc cc cc}
\toprule
\multirow{3}{*}{\textbf{Backbone}} &
\multirow{3}{*}{\textbf{Feature}} &
\multicolumn{2}{c}{\textbf{HCP}} &
\multicolumn{2}{c}{\textbf{PPMI}} &
\multicolumn{2}{c}{\textbf{ADNI (MCI)}} &
\multicolumn{2}{c}{\textbf{ADNI (AD)}} \\

& & \multicolumn{2}{c}{\textit{Sex Classif.}} &
\multicolumn{2}{c}{\textit{PD Diagnosis}} &
\multicolumn{2}{c}{\textit{Diagnosis}} &
\multicolumn{2}{c}{\textit{Diagnosis}} \\

\cmidrule(lr){3-4} \cmidrule(lr){5-6}\cmidrule(lr){7-8}\cmidrule(lr){9-10}
& & \textbf{ACC $\uparrow$} & \textbf{wF1 $\uparrow$}
& \textbf{ACC $\uparrow$} & \textbf{wF1 $\uparrow$}
& \textbf{ACC $\uparrow$} & \textbf{wF1 $\uparrow$}
& \textbf{ACC $\uparrow$} & \textbf{wF1 $\uparrow$} \\
\midrule

DINOv2 & One global token & 80.37$\pm$2.55 & 80.33$\pm$2.53 & 51.39$\pm$3.18 & 51.58$\pm$2.25 & 60.68$\pm$2.67 & 60.44$\pm$1.98 & 75.46$\pm$4.01 & 75.20$\pm$3.69 \\
DINOv2 & 40 Global tokens        & 79.53$\pm$1.17 & 79.50$\pm$1.15 & 47.57$\pm$5.14 & 48.82$\pm$2.67 & 55.98$\pm$2.67 & 55.58$\pm$2.09 & 75.00$\pm$2.41 & 74.61$\pm$2.34 \\
\midrule

ViT-B & 40 Global tokens         & 70.00$\pm$1.13 & 70.02$\pm$1.15 & 60.00$\pm$1.36 & 55.45$\pm$1.72 & 52.76$\pm$4.57 & 52.19$\pm$4.71 & 78.11$\pm$2.56 & 76.84$\pm$2.71 \\
\midrule
\rowcolor{rowgray}
SigLIP2 & One Global token        & 82.06$\pm$2.05 & 82.04$\pm$2.05 & 54.86$\pm$1.96 & 55.72$\pm$1.63 & 61.11$\pm$1.48 & 60.75$\pm$1.16 & 75.46$\pm$2.12 & 75.41$\pm$2.11 \\
SigLIP2 & 40 Global tokens         & 83.08$\pm$0.59 & 83.02$\pm$0.66 & 53.47$\pm$7.54 & 53.93$\pm$6.14 & 60.68$\pm$1.96 & 60.02$\pm$2.07 & 74.07$\pm$4.88 & 74.13$\pm$4.98 \\
SigLIP2 & 40 patch\_mean & 84.94$\pm$1.06 & 84.91$\pm$1.06 & 56.94$\pm$4.21 & 56.48$\pm$4.27 & 58.12$\pm$1.96 & 57.82$\pm$1.87 & 76.39$\pm$4.17 & 76.36$\pm$3.93 \\

\bottomrule
\end{tabular}
}
\end{table*}

\subsection{Configuration}
\label{app:our_parameters}

All FlatClip encoders are frozen during downstream evaluation. 
We first render scalar cortical fMRI activity into RGB cortical flatmaps and then extract image features with a frozen SigLIP2-NaFlex encoder. 
Only the downstream MLP probe is trained for task prediction. 
The main settings are summarized in Table~\ref{tab:flatclip_settings}. We choose $T = 40$, following the setting with prior works~\cite{Wang2025TowardsAG,wang2026omni}.

\begin{table}[!h]
\centering
\small
\caption{Hyperparameter settings of FlatClip.}
\label{tab:flatclip_settings}
\begin{tabular}{ll}
\toprule
\textit{Configuration} & \textit{Value} \\
\midrule
\multicolumn{2}{c}{\textbf{Flatmap rendering}} \\
\midrule
surface space & fsLR cortical surface \\
renderer & PyCortex / cached flatmap projection \\
Normalization & global z-score with bounded color scale \\
colormap & Inferno (HCP); RdBu\_r (NSD) \\
background pixels & transparent outside cortex/ROI, composited on white RGB \\
flatmap crop & cropped to non-background cortical pixels \\
\midrule
\multicolumn{2}{c}{\textbf{Resting-state configs}} \\
\midrule
image encoder & frozen SigLIP2-base-patch16-NaFlex \\
input format & sequence of cortical flatmap frames \\
input region & whole cortex \\
frames per subject & first 40 resting-state frames \\
encoder input & RGB flatmap, padded/processed by SigLIP2-NaFlex processor \\
frame feature & global image embedding \\
frame feature dimension & 768 \\
subject feature & mean pooling over 40 frame features \\
probe architecture & two hidden layers: $768\to256\to256\to C$, dropout 0.2 \\
optimizer & AdamW \\
learning rate & $1\times10^{-3}$ \\
weight decay & $1\times10^{-4}$ \\
batch size & 64 \\
training epochs & up to 250, early stopping patience 40 \\
validation metric & weighted-F1 \\
loss function & class-weighted cross-entropy \\
reported metrics & accuracy, weighted-F1 \\
\midrule
\multicolumn{2}{c}{\textbf{NSD visual-fMRI configs}} \\
\midrule
image encoder & frozen SigLIP2-base-patch16-NaFlex \\
input format & stimulus-level GLM cortical flatmap \\
input regions & whole cortex / HCP-MMP visual / NSDgeneral \\
encoder output & mean over 256 adaptively pooled feature tokens \\
main feature dimension & 768 \\
probe architecture & MLP with LayerNorm, hidden dim 1024, dropout 0.1 \\
optimizer & AdamW \\
learning rate & $3\times10^{-4}$ \\
weight decay & $1\times10^{-4}$ \\
batch size & 128 \\
training epochs & 10 \\
validation metric & mAP \\
loss function & class-weighted BCE with logits \\
reported metrics & mAP, weighted-F1 \\
\bottomrule
\end{tabular}
\end{table}

\paragraph{Normalization and inference cost.}
Following the rendering steps in Section~\ref{sec:flatmap_adapter}, resting-state inputs are globally z-scored over the selected vertex-by-time matrix. NSD responses are z-scored within each GLM map and mapped to RGB using a diverging colormap. Native crop dimensions are listed in Table~\ref{tab:added_dimensions}; SigLIP2-NaFlex converts these variable-size images into a fixed-budget patch sequence.

Feature extraction is performed offline with the frozen SigLIP2-NaFlex encoder. Image counts below refer to extraction jobs; downstream cohort sizes refer to matched feature and label sets.
From the logged extraction jobs, HCP required 3.96 hours on four A100 GPUs for 40,440 flatmap images, corresponding to 2.84 images/s in wall-clock throughput. 
PPMI required 39.5 minutes on four A100 GPUs for 18,960 images, and ADNI required 61.6 minutes on two A100 GPUs for 19,840 images. 
After features are cached, downstream MLP training is lightweight and does not require backpropagation through the image encoder.

\begin{table*}[t]
\centering
\small
\setlength{\tabcolsep}{3pt}
\renewcommand{\arraystretch}{1.2}
\caption{Within-subject COCO80 multi-label classification on NSD. 
\best{Red} indicates the best performance. wF1 denotes weighted-F1. 
Unmarked FlatClip rows use the main setting: one 768-dimensional feature obtained by averaging a $16\times16$ feature grid adaptively pooled from valid NaFlex output tokens. This grid is formed after encoding and is distinct from the processor's variable-size patch grid.
Rows marked with $^{*}$ retain all 256 pooled feature tokens.}
\label{tab:nsd_coco80_within_map_wf1}

\resizebox{\textwidth}{!}{
\begin{tabular}{l cc cc cc cc cc}
\toprule
\multirow{2}{*}{\diagbox[width=7em]{\textbf{Model}}{\textbf{Subject}}} &
\multicolumn{2}{c}{\textbf{Sub1}} &
\multicolumn{2}{c}{\textbf{Sub2}} &
\multicolumn{2}{c}{\textbf{Sub5}} &
\multicolumn{2}{c}{\textbf{Sub7}} &
\multicolumn{2}{c}{\textbf{Mean}} \\

\cmidrule(lr){2-3} \cmidrule(lr){4-5} \cmidrule(lr){6-7} \cmidrule(lr){8-9} \cmidrule(lr){10-11}
& \textbf{mAP $\uparrow$} & \textbf{wF1 $\uparrow$}
& \textbf{mAP $\uparrow$} & \textbf{wF1 $\uparrow$}
& \textbf{mAP $\uparrow$} & \textbf{wF1 $\uparrow$}
& \textbf{mAP $\uparrow$} & \textbf{wF1 $\uparrow$}
& \textbf{mAP $\uparrow$} & \textbf{wF1 $\uparrow$} \\
\midrule

SigLIP2-NSDgeneral & \best{15.311} & \best{26.922} & 15.714 & 26.986 & \best{19.933} & \best{29.611} & \best{14.428} & 24.366 & \best{16.347} & \best{26.971} \\
SigLIP2-Visual     & 14.050 & 25.829 & \best{16.065} & \best{27.151} & 18.889 & 29.476 & 14.361 & \best{25.238} & 15.841 & 26.924 \\
DINOv2-Visual      & 14.765 & 26.338 & 14.780 & 25.105 & 17.213 & 27.304 & 12.605 & 22.547 & 14.841 & 25.323 \\
ViT-Cortex               & 11.674 & 23.092 & 13.373 & 24.128 & 13.895 & 26.376 & 10.544 & 20.432 & 12.371 & 23.507 \\
SigLIP2-Cortex     & 11.100 & 24.576 & 12.211 & 25.048 & 14.974 & 26.090 & 10.053 & 18.822 & 12.085 & 23.634 \\
DINOv2-Cortex      & 10.930 & 22.800 & 10.815 & 23.022 & 11.290 & 24.427 & 9.467 & 19.751 & 10.625 & 22.500 \\
SwiFT             & 7.974 & 20.582 & 10.769 & 22.303 & 10.274 & 22.989 & 8.478 & 20.858 & 9.374 & 21.683 \\
NeuroSTORM         & 6.491 & 12.358 & 8.798 & 20.484 & 9.253 & 10.471 & 8.209 & 18.326 & 8.188 & 15.410 \\
Omni-fMRI          & 3.949 & 13.818 & 5.024 & 15.910 & 4.715 & 13.441 & 3.751 & 8.934 & 4.360 & 13.026 \\

\midrule
SigLIP2-Cortex$^{*}$ & 16.604 & 29.776 & 15.771 & 28.262 & 19.218 & 31.379 & 14.129 & 27.836 & 16.430 & 29.313 \\
SigLIP2-Visual$^{*}$ & 20.192 & \best{33.255} & 19.623 & 29.727 & 23.323 & \best{33.801} & 19.490 & 30.457 & 20.657 & 31.810 \\
SigLIP2-NSDgeneral$^{*}$ & \best{21.565} & 32.330 & \best{22.218} & \best{32.084} & \best{25.300} & 32.744 & \best{20.282} & \best{31.123} & \best{22.341} & \best{32.070} \\

\bottomrule
\end{tabular}
}
\end{table*}

\subsection{The influence of frame number}

We further evaluate the effect of temporal sampling length in resting-state FlatClip.
For each subject, we uniformly sample 10, 20, or 40 frames from the available 40-frame sequence.
Each frame is encoded independently by the frozen SigLIP2 encoder, and the selected frame-level global representations are mean-pooled into a single 768-dimensional subject-level feature.
As shown in Table~\ref{tab:rest_frame_count_ablation}, increasing the number of frames generally improves or stabilizes performance, but the gain is modest under the one-token pooling strategy.
HCP and ADNI show the clearest benefit from using more temporal samples, with 40 frames giving the best or tied-best performance in most metrics.
ADNI (AD) already saturates around 20 frames, while PPMI does not show a consistent improvement with additional frames, suggesting that temporal coverage alone is not the main limiting factor for this task.
Overall, these results indicate that FlatClip can extract useful subject-level information even from sparse temporal samples, while the 40-frame setting provides a more stable default for the main experiments.

\begin{table*}[t]
\centering
\small
\setlength{\tabcolsep}{2pt}
\renewcommand{\arraystretch}{1.2}
\caption{Ablation on the number of resting-state frames used by FlatClip.
All results use frozen SigLIP2 with one global representation per frame.
Selected frame features are mean-pooled into a 768-dimensional subject-level token.
Results are reported as Accuracy / weighted F1 over three seeds.}
\label{tab:rest_frame_count_ablation}

\resizebox{\textwidth}{!}{
\begin{tabular}{l cc cc cc cc}
\toprule
\multirow{3}{*}{\diagbox[width=6em]{\textbf{Frames}}{\textbf{Dataset}}} &
\multicolumn{2}{c}{\textbf{HCP}} &
\multicolumn{2}{c}{\textbf{ADNI (MCI)}} &
\multicolumn{2}{c}{\textbf{ADNI (AD)}} &
\multicolumn{2}{c}{\textbf{PPMI}} \\

& \multicolumn{2}{c}{\textit{Sex Classif.}} &
\multicolumn{2}{c}{\textit{Diagnosis}} &
\multicolumn{2}{c}{\textit{Diagnosis}} &
\multicolumn{2}{c}{\textit{PD Diagnosis}} \\

\cmidrule(lr){2-3} \cmidrule(lr){4-5}\cmidrule(lr){6-7}\cmidrule(lr){8-9}
& \textbf{ACC $\uparrow$} & \textbf{wF1 $\uparrow$}
& \textbf{ACC $\uparrow$} & \textbf{wF1 $\uparrow$}
& \textbf{ACC $\uparrow$} & \textbf{wF1 $\uparrow$}
& \textbf{ACC $\uparrow$} & \textbf{wF1 $\uparrow$} \\
\midrule

10 frames
& 76.31$\pm$4.61 & 76.27$\pm$4.64
& 59.40$\pm$3.23 & 58.52$\pm$3.13
& 76.39$\pm$1.39 & 76.39$\pm$1.43
& 54.17$\pm$2.08 & 54.12$\pm$2.11 \\

20 frames
& 77.83$\pm$1.92 & 77.79$\pm$1.86
& 60.68$\pm$1.48 & 59.90$\pm$1.77
& \best{78.24$\pm$0.80} & \best{78.29$\pm$1.26}
& 52.43$\pm$1.59 & 53.39$\pm$2.27 \\

\rowcolor{rowgray}
40 frames
& \best{82.06$\pm$2.05} & \best{82.04$\pm$2.05}
& \best{61.11$\pm$1.48} & \best{60.75$\pm$1.16}
& 75.46$\pm$2.12 & 75.41$\pm$2.11
& \best{54.86$\pm$1.96} & \best{55.72$\pm$1.63} \\

\bottomrule
\end{tabular}
}
\end{table*}

\section{Details of NSD Visual-fMRI Experiments}
\label{sec:supp_ablation}

\paragraph{Natural Scenes Dataset (NSD).}
We evaluate visual-fMRI decoding on the Natural Scenes Dataset (NSD), a large-scale 7T fMRI dataset in which subjects viewed natural images from COCO. 
We use stimulus-level GLM beta estimates, specifically the \texttt{betas\_fithrf\_GLMdenoise\_RR} version, as neural response inputs. 
The decoding target is an 80-dimensional COCO multi-hot label vector.
For each evaluated subject, 8,550 non-shared images are used for training, 450 non-shared images for validation, and the 1,000 shared images for testing. Image identities are disjoint across these partitions.

\subsection{GLM for Baselines models}

\textbf{SwiFT} extracted features in a volume-wise manner: each NSD beta volume is processed independently, and the resulting representation is used as the feature for that volume.

\textbf{NeuroSTORM} follows a similar strategy. Its NSD extraction pipeline also processes each beta volume independently with a 3D backbone and saves the resulting backbone feature for each volume.

\textbf{Omni-fMRI} expects 40 input channels. For the stimulus-level beta-map baseline, a single 3D beta volume is repeated along this dimension and encoded into a CLS feature for downstream prediction.
The native-BOLD experiment in Appendix~\ref{app:added_omni} additionally evaluates temporally varying inputs with the same released encoder.

\subsection{Supplementary Analysis of Image--fMRI Feature Alignment}
\label{sec:supp_image_fmri_alignment}

\begin{figure*}[t]
\centering
\includegraphics[width=\textwidth]{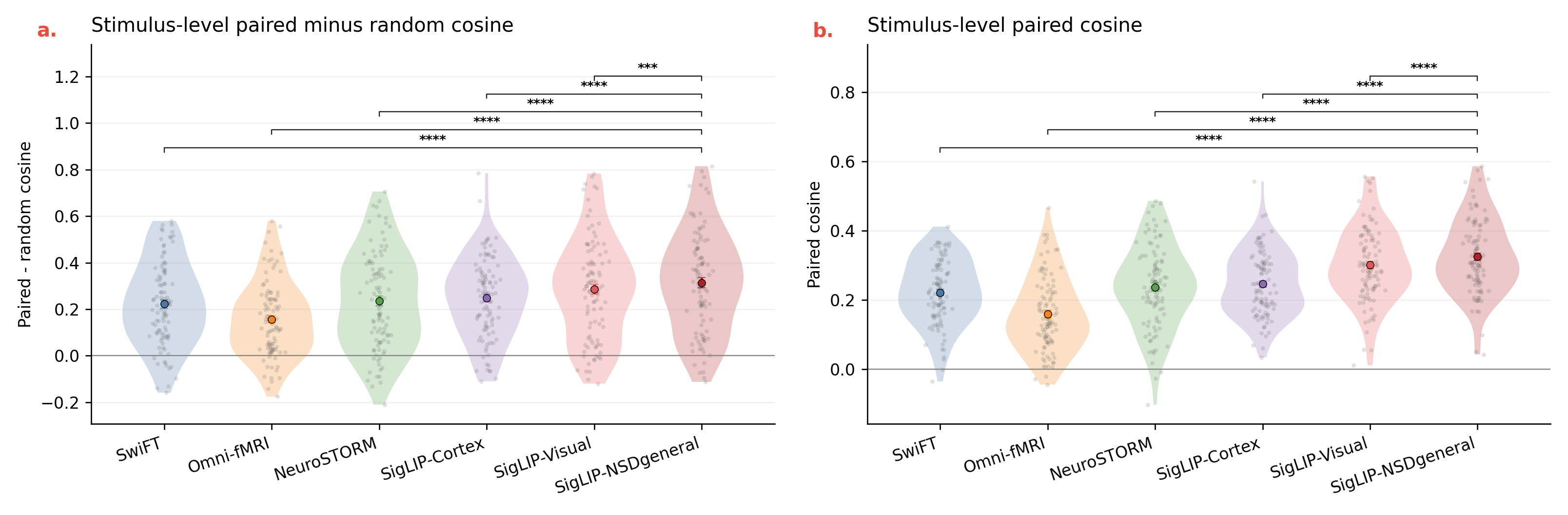}
\caption{Stimulus-level image--fMRI feature alignment analysis on a randomly sampled subset of NSD shared1000 in SigLIP2 space.
(a) Paired-random cosine margin, defined as the cosine similarity between the aligned fMRI-derived feature and the matched image feature minus a randomly mismatched image cosine.
(b) Paired cosine similarity between the aligned fMRI-derived feature and its matched image feature.
For each shared image, scores are first computed for each subject and then averaged across the four subjects.
Dots denote individual shared stimuli, and large markers denote the mean.
Significance brackets compare SigLIP2-NSDgeneral against each other method using paired Wilcoxon signed-rank tests over stimuli. Cosines are computed after separately fitted subject- and model-specific ridge alignment.}
\label{fig:supp_siglip_image_fmri_alignment}
\end{figure*}

\begin{table*}[t]
\centering
\small
\setlength{\tabcolsep}{3pt}
\renewcommand{\arraystretch}{1.12}
\caption{Stimulus-level image--fMRI feature alignment analysis on a randomly sampled subset of NSD shared1000 in SigLIP2 space.
Values are averaged over four subjects, leaving one value per image.
$\Delta$ denotes SigLIP2-NSDgeneral minus the comparison method.
$p_t$ and $p_w$ denote one-sided paired $t$-test and Wilcoxon signed-rank p-values, with Holm correction across the 15 comparisons within each test family. Confidence intervals are percentile bootstrap intervals over stimulus-level paired differences.}
\label{tab:supp_siglip_image_fmri_alignment}

\resizebox{\textwidth}{!}{
\begin{tabular}{l l c c c c c c c}
\toprule
\textbf{Metric} & \textbf{Comparison} & \textbf{Baseline} & \textbf{SigLIP2-NSDgeneral} & \textbf{$\Delta$} & \textbf{95\% CI} & \textbf{$p_t$} & \textbf{$p_w$} & \textbf{Improved} \\
\midrule
Local-RDM & vs SwiFT          & 0.026 & 0.134 & 0.108 & [0.086, 0.129] & $9.01{\times}10^{-16}$ & $3.18{\times}10^{-12}$ & 83.0\% \\
Local-RDM & vs Omni-fMRI      & 0.009 & 0.134 & 0.125 & [0.102, 0.147] & $2.70{\times}10^{-17}$ & $5.27{\times}10^{-13}$ & 86.0\% \\
Local-RDM & vs NeuroSTORM     & 0.052 & 0.134 & 0.083 & [0.064, 0.101] & $5.78{\times}10^{-13}$ & $1.34{\times}10^{-10}$ & 79.0\% \\
Local-RDM & vs SigLIP2-Cortex  & 0.032 & 0.134 & 0.102 & [0.084, 0.121] & $2.09{\times}10^{-17}$ & $2.87{\times}10^{-13}$ & 89.0\% \\
Local-RDM & vs SigLIP2-Visual  & 0.091 & 0.134 & 0.043 & [0.026, 0.061] & $5.41{\times}10^{-6}$  & $5.73{\times}10^{-6}$  & 74.0\% \\
\midrule
Paired cosine & vs SwiFT          & 0.222 & 0.325 & 0.103 & [0.087, 0.119] & $3.82{\times}10^{-21}$ & $7.31{\times}10^{-15}$ & 90.0\% \\
Paired cosine & vs Omni-fMRI      & 0.159 & 0.325 & 0.166 & [0.150, 0.182] & $1.50{\times}10^{-35}$ & $3.61{\times}10^{-17}$ & 99.0\% \\
Paired cosine & vs NeuroSTORM     & 0.237 & 0.325 & 0.088 & [0.072, 0.104] & $5.07{\times}10^{-17}$ & $4.31{\times}10^{-14}$ & 89.0\% \\
Paired cosine & vs SigLIP2-Cortex  & 0.246 & 0.325 & 0.079 & [0.064, 0.094] & $3.70{\times}10^{-16}$ & $1.17{\times}10^{-13}$ & 89.0\% \\
Paired cosine & vs SigLIP2-Visual  & 0.301 & 0.325 & 0.024 & [0.015, 0.033] & $3.83{\times}10^{-6}$  & $7.96{\times}10^{-6}$  & 69.0\% \\
\midrule
Margin & vs SwiFT          & 0.223 & 0.314 & 0.091 & [0.066, 0.116] & $7.19{\times}10^{-10}$ & $8.30{\times}10^{-9}$  & 76.0\% \\
Margin & vs Omni-fMRI      & 0.156 & 0.314 & 0.157 & [0.128, 0.186] & $5.07{\times}10^{-17}$ & $2.87{\times}10^{-13}$ & 87.0\% \\
Margin & vs NeuroSTORM     & 0.235 & 0.314 & 0.079 & [0.053, 0.106] & $1.81{\times}10^{-7}$  & $4.49{\times}10^{-7}$  & 72.0\% \\
Margin & vs SigLIP2-Cortex  & 0.248 & 0.314 & 0.066 & [0.042, 0.090] & $1.38{\times}10^{-6}$  & $2.49{\times}10^{-6}$  & 69.0\% \\
Margin & vs SigLIP2-Visual  & 0.286 & 0.314 & 0.028 & [0.012, 0.044] & $4.23{\times}10^{-4}$  & $4.06{\times}10^{-4}$  & 68.0\% \\
\bottomrule
\end{tabular}
}
\end{table*}

We further quantified image--fMRI feature alignment in SigLIP2 space using a randomly sampled 100-image subset from NSD shared1000
(Fig.~\ref{fig:supp_siglip_image_fmri_alignment}, Table~\ref{tab:supp_siglip_image_fmri_alignment}).
For each shared image, we computed the alignment score for each subject and then averaged across the four subjects, yielding one value per image.
For paired cosine and margin, a separate ridge map is fitted for each subject and fMRI backbone to predict SigLIP2 image features. The evaluation set consists of 100 shared images sampled with seed 42; mapping uses the other available matched images, including shared images outside this held-out subset. Feature standardization and PCA are fitted on the mapping-training set only. The fMRI features are reduced to at most 256 PCA components, followed by ridge regression with $\alpha=100$ and a fitted intercept. These hyperparameters are fixed rather than selected on the held-out images. Cosine similarity is computed between the ridge predictions and the matched image features standardized with the training-set statistics. Local-RDM is instead computed directly from distance profiles in each model's own feature space, without the learned mapping.
For Local-RDM, each image was represented by its cosine-distance profile to the other evaluation images after row-wise standardization and L2 normalization, and we computed the Spearman correlation with the corresponding SigLIP2 image-feature distance profile.
Paired cosine measured the similarity between the aligned fMRI-derived feature and its matched image feature, while the paired-random margin subtracted a randomly mismatched image cosine.

Across all three stimulus-level metrics, SigLIP2-NSDgeneral achieved the strongest alignment with the corresponding image features.
Compared with the strongest non-FlatClip baseline, NeuroSTORM, SigLIP2-NSDgeneral improved Local-RDM by 0.083, paired cosine by 0.088, and paired-random margin by 0.079.
The advantage also remained significant when compared with SigLIP2-Visual, indicating that restricting the flatmap input to NSD task-active cortex yields representations that are more tightly aligned with image-feature geometry than using a broader visual mask.
This analysis complements COCO80 classification by quantifying stimulus-level correspondence between cortical and image representations.

\subsection{Ablation and Detailed Results}

Table~\ref{tab:nsd_coco80_within_map_wf1} gives subject-wise results for the frozen image backbones and cortical regions. Unmarked rows use one pooled representation; starred rows retain all 256 adaptively pooled feature tokens to assess sensitivity to the feature representation.

\section{Details of Geometry perturbation Experiments}
\label{app:perturbation}

\subsection{Geometry-control perturbations.}
We constructed spatial-perturbation controls for HCP sex classification to assess the contribution of cortical image layout.
All input controls used the same subjects, train/validation/test split, number of frames, frozen pretrained encoder, and MLP architecture and training protocol; the random-initialization condition replaces only the encoder weights. Each condition used an independently trained probe selected on validation data.
For each subject, we used a fixed sequence of 40 frames. 
All rendered inputs were converted to the same image canvas before feature extraction, so that performance differences primarily reflect changes in spatial organization under matched downstream settings, while keeping image canvas handling and feature extraction fixed within each control family.

\textbf{Real flatmap.}
The real-flatmap condition uses the original cortical flatmap rendering without spatial perturbation. 
Each frame preserves the native 2D cortical surface layout produced by the flatmap projection, including both local activity patterns and their global anatomical arrangement.

\textbf{Left-right hemisphere swap.}
For the left-right swap control, we exchanged the left and right hemispheric halves of each rendered flatmap image. 
The same transformation was applied to all subjects and frames. 
This control preserves the pixel intensity distribution, local texture, and most low-level image statistics, but disrupts hemispheric laterality and the correspondence between cortical location and visual image position. 
It tests sensitivity to the placement of the two hemispheres within the image.

\textbf{Within-hemisphere region shuffle.}
For the within-hemisphere shuffle control, we first separated the rendered flatmap into left- and right-hemisphere image regions. 
Within each hemisphere, the valid cortical area was divided into local image blocks, and these blocks were randomly permuted only within the same hemisphere. 
The same block permutation was applied consistently across subjects and frames. 
This preserves hemispheric identity and local block-level activation patterns, while disrupting the finer global layout of cortical regions within each hemisphere. 
Compared with left-right swapping, this provides a stronger test of whether within-hemisphere cortical organization contributes to downstream prediction.

\textbf{Phase-randomized, histogram-matched.}
To further control for rendering-induced low-level image statistics, we constructed a phase-randomized control for each flatmap frame. 
For each RGB channel, we fill the non-cortical background with the foreground mean, retain the resulting Fourier amplitudes, randomize phases, and transform back to image space. Foreground values are then rank-matched to the original foreground intensity histogram, and the white background is restored. This preserves the cortical silhouette and foreground intensity distribution while disrupting spatial organization; the subsequent histogram matching can change the Fourier amplitudes, so frequency matching is approximate.

\textbf{Random initialized encoder.}
In addition to input perturbations, we included a randomly initialized SigLIP2 encoder control. 
This condition uses the real flatmap input and the same downstream MLP probe, but removes pretrained visual representations by replacing the frozen pretrained encoder with the same architecture initialized from random weights. 
This control tests whether the observed performance comes only from the model architecture and image-format input, or whether pretrained visual features contribute to the FlatClip representation.

\textbf{Spatial block-shuffle.}
For the block-shuffle control, each flatmap image was divided into non-overlapping $16\times16$ image blocks. 
A single random block permutation was generated and applied consistently to all subjects and all frames. 
This preserves the local texture and activation pattern inside each block, but disrupts the global arrangement of cortical regions. 
Thus, this control tests whether local patch-level fMRI patterns alone are sufficient when large-scale cortical topology is scrambled.

\textbf{Cortical-pixel permutation.}
For the cortical-pixel permutation control, we identified valid cortical pixels using the flatmap alpha mask and randomly permuted the values within this cortical mask. 
The same permutation was applied consistently across all subjects and frames. 
This preserves the rendered intensity distribution within each frame and leaves the background unchanged, while disrupting local image neighborhoods and the correspondence between cortical location and signal values.
Therefore, this is a stronger geometry disruption than block-shuffling.

\textbf{FC heatmap.}
For the FC heatmap control, we replaced frame-wise flatmaps with a subject-level functional connectivity image. 
For each subject, one 200-frame Schaefer100 ROI time series was used to compute a $100\times100$ Pearson correlation matrix after ROI-wise normalization. 
The resulting FC matrix was clipped to $[-1,1]$, rendered as a red--blue heatmap, resized to the same image canvas, and repeated as 40 identical frames. 
This control keeps subject-level functional information but removes cortical surface geometry and frame-wise flatmap dynamics.

\textbf{Random image.}
For the random-image control, each frame was replaced by a random RGB image with the same spatial size as the flatmap input. 
Random images were generated with deterministic subject- and frame-specific seeds. 
This control preserves only the image-format interface to the frozen visual encoder, while removing both fMRI signal and cortical geometry.

As an additional robustness check, we repeat the geometry-control ablation with a frozen DINOv2 encoder. 
Table~\ref{tab:hcp_geometry_control_dino} shows the same qualitative trend as the main SigLIP2 geometry-control experiment.

\begin{table}[t]
\centering
\tiny
\setlength{\tabcolsep}{4pt}
\caption{Geometry-control ablation on HCP sex classification.
All results use the frozen DINOv2 encoder with one global token per frame, followed by an independently trained MLP with the same architecture and training protocol for every condition.
Results are reported as mean$\pm$std over three seeds.
Significance markers indicate paired $t$-tests over the three probe-seed results, comparing Real flatmap against each control after Holm correction.
\best{Red} indicates the best performance.}
\label{tab:hcp_geometry_control_dino}

\resizebox{\linewidth}{!}{
\begin{tabular}{l llll}
\toprule
\textbf{Control input} &
\textbf{ACC $\uparrow$} &
\textbf{wF1 $\uparrow$} &
\textbf{Balanced ACC $\uparrow$} &
\textbf{Macro F1 $\uparrow$} \\
\midrule
\rowcolor{rowgray}
Real flatmap
& \best{80.37\,${\pm}$\,2.55}
& \best{80.33\,${\pm}$\,2.53}
& \best{80.08\,${\pm}$\,2.47}
& \best{80.15\,${\pm}$\,2.53} \\

Spatial block-shuffle
& 75.47\,${\pm}$\,0.78$^{*}$
& 75.49\,${\pm}$\,0.76$^{*}$
& 75.44\,${\pm}$\,0.70$^{*}$
& 75.34\,${\pm}$\,0.75$^{*}$ \\

Cortical-pixel permutation
& 73.77\,${\pm}$\,2.80$^{*}$
& 73.80\,${\pm}$\,2.81$^{*}$
& 74.06\,${\pm}$\,2.61$^{*}$
& 73.73\,${\pm}$\,2.76$^{*}$ \\

FC heatmap
& 54.99\,${\pm}$\,5.08$^{**}$
& 55.04\,${\pm}$\,5.07$^{**}$
& 54.80\,${\pm}$\,5.08$^{**}$
& 54.77\,${\pm}$\,5.07$^{**}$ \\

Random image
& 49.24\,${\pm}$\,1.52$^{**}$
& 49.33\,${\pm}$\,1.52$^{**}$
& 49.24\,${\pm}$\,1.54$^{**}$
& 49.14\,${\pm}$\,1.53$^{**}$ \\
\bottomrule
\end{tabular}
}

\vspace{2pt}
\footnotesize{
Markers denote Holm-corrected paired $t$-tests against Real flatmap:
$^{*}p<0.05$, $^{**}p<0.01$.
}
\end{table}

\subsection{Details of ROI and voxel reference models}
\label{sec:supp_roi_voxel}

Table~\ref{tab:representation_comparison} compares a regional time-series model with native voxel models using standard or adaptive patching. The models use subject-level train/validation/test partitions and 40-frame inputs.

\paragraph{Native voxel-volume processing.}
For the voxel-volume setting, we use preprocessed 4D fMRI scans and a 40-frame sampling strategy, retaining voxel-wise signals.
Voxels outside the selected brain or ROI mask are set to zero, and temporal z-score normalization is applied voxel-wise. 
This setting keeps fine-grained voxel-level spatial variation and serves as a stronger geometry-preserving volumetric reference.

\paragraph{Voxel models and patching.}
The voxel models use a ViT-Small backbone. We compare standard patching with adaptive patch allocation following Omni-fMRI~\cite{wang2026omni}.

\paragraph{40-frame ROI time-series processing.}
For the ROI time-series setting, we convert each 4D resting-state scan into a region-level temporal representation using the same ROI definition. 
At each time point, voxel signals within each ROI are averaged to obtain a regional activity vector. 
For subject $i$, this produces a time series
\[
R_i = [r_{i,1}, r_{i,2}, \ldots, r_{i,T}], \quad r_{i,t} \in \mathbb{R}^{K},
\]
where $K$ is the number of ROIs. 
The regional signals are temporally z-scored within each ROI, and the same 40-frame sampling strategy is applied to obtain
\[
R_i^{40} \in \mathbb{R}^{40 \times K}.
\]
This representation removes within-ROI spatial geometry and keeps only region-level temporal dynamics. 
We feed the 40-frame ROI sequence to a lightweight temporal model using the same subject-level train/validation/test partitions as the voxel reference models.

All experiments are performed at the subject level. 
Frames from the same subject are never split across train, validation, and test sets, preventing frame-level leakage.

\section{Additional Evaluation Protocols and Results}
\label{app:added_results}

\subsection{Native Image Dimensions}
\label{app:added_dimensions}

Table~\ref{tab:added_dimensions} lists the native width$\times$height of the RGB flatmaps constructed in Section~\ref{sec:flatmap_adapter}, before SigLIP2-NaFlex processing.

\begin{table}[ht]
\centering\small
\caption{Native cropped flatmap dimensions before encoder processing.}
\label{tab:added_dimensions}
\begin{tabular}{lc}
\toprule
Input region & Width$\times$height \\
\midrule
Whole cortex & $943\times384$ \\
HCP-MMP visual & $273\times264$ \\
NSDgeneral, subject 1 & $293\times208$ \\
NSDgeneral, subjects 2 and 5 & $293\times209$ \\
NSDgeneral, subject 7 & $292\times209$ \\
\bottomrule
\end{tabular}
\end{table}

\subsection{Surface-Model Implementations}
\label{app:added_surface}

\paragraph{CortexMAE-F.}
We use the released CortexMAE-F checkpoint and its standard surface preprocessing pipeline, pooling frozen backbone features into a subject-level representation for the downstream probe. Pretraining includes HCP-YA; the ADNI tasks evaluate transfer to a different cohort.

\paragraph{SiT-style parcel-token control.}
The SiT-style results in Table~\ref{tab:added_surface_models} use cortical time series read from fsLR CIFTI dtseries files. Schaefer1000 parcel averages are grouped into 320 connected surface patches, which are processed by a Transformer with width 192, depth 12, and three attention heads. The reported configuration uses learnable positional embeddings and is trained end to end with seeds 42, 43, and 44. It operates on surface signals without color mapping or image rendering.

\paragraph{Vertex-level spherical patching.}
For the additional vertex-level implementation, GIFTI spherical surfaces provide the geometry for resampling fsLR signals to ico6. We then use the official SiT ico2 indices, with 320 patches per hemisphere and 153 ico6 vertices per patch. In resting-state inputs, the 40-frame values within each patch are projected to a 192-dimensional token. The two hemispheres share the Transformer, and their logits are averaged at evaluation. For NSD, beta responses are used in place of temporal windows, and the visual or NSDgeneral mask selects the input support.

\subsection{Measured-Confound Sensitivity}
\label{app:added_confounds}

We fit covariate encodings and regression models on the training split only. Let $H$ denote the frozen feature matrix and $A$ the encoded covariates, including an intercept. We estimate a regression coefficient matrix $B$ from $(A_{\mathrm{train}},H_{\mathrm{train}})$ and construct adjusted features $H^{\mathrm{adj}}=H-AB$ using the same fitted coefficients for validation and test subjects. We then standardize the features using training statistics and train the classification probe. The diagnostic labels remain the classification targets.

Table~\ref{tab:added_confounds} reports each covariate adjustment alongside its task-specific raw reference. The PPMI raw row is the re-evaluation paired with its covariate analyses.

\begin{table}[ht]
\centering\small
\caption{\textbf{Measured-confound sensitivity.} Mean$\pm$std (\%) over three probe seeds. $\Delta$wF1 is the percentage-point change from the raw row for the same task.}
\label{tab:added_confounds}
\resizebox{\linewidth}{!}{\begin{tabular}{llrrr}
\toprule
Task & Adjustment & ACC & wF1 & $\Delta$wF1 \\
\midrule
HCP sex & Raw & $82.06\pm2.05$ & $82.04\pm2.05$ & Ref. \\
 & Age & $80.88\pm1.06$ & $80.88\pm1.07$ & $-1.16$ \\
\midrule
ADNI MCI/CN & Raw & $61.11\pm1.48$ & $60.75\pm1.16$ & Ref. \\
 & Age + sex & $59.40\pm0.74$ & $59.03\pm0.38$ & $-1.72$ \\
 & Age + sex + site & $57.69\pm0.00$ & $56.47\pm0.60$ & $-4.28$ \\
 & Age + sex + site + protocol & $58.12\pm1.96$ & $57.69\pm1.57$ & $-3.06$ \\
\midrule
ADNI AD/CN & Raw & $75.46\pm2.12$ & $75.41\pm2.11$ & Ref. \\
 & Motion & $75.00\pm1.39$ & $75.34\pm1.56$ & $-0.07$ \\
 & Age + sex + motion & $69.44\pm4.81$ & $69.96\pm4.73$ & $-5.45$ \\
\midrule
PPMI three-way & Raw re-evaluation & $54.51\pm5.35$ & $55.59\pm4.20$ & Ref. \\
 & Age + sex & $57.29\pm3.13$ & $57.49\pm3.62$ & $+1.90$ \\
 & Age + sex + acquisition/cohort & $53.47\pm3.35$ & $52.52\pm3.19$ & $-3.07$ \\
\bottomrule
\end{tabular}}
\end{table}

For the MCI/CN age+sex+site condition, all three seeds yield the same number of correct test predictions, producing zero standard deviation in accuracy. Their error patterns differ, producing a nonzero standard deviation in weighted F1. Site identifiers and protocol descriptions are encoded as their respective metadata fields. Signal-quality proxies, when used in the volume controls, refer to volume-derived DVARS and global-signal summaries.

\subsection{Alternative Forty-Frame Windows}
\label{app:added_windows}

Table~\ref{tab:added_windows} compares TR0--39 and TR98--137 using the same encoder and probe protocol. MCI/CN is re-evaluated on both windows. The alternative AD/CN window uses 243/34/72 training/validation/test subjects: one training acquisition contains only 97 TRs, while validation and test membership is unchanged.

\begin{table}[ht]
\centering\small
\caption{\textbf{Alternative 40-TR windows.} Accuracy and weighted F1 (\%). The first-window MCI/CN row is the paired re-evaluation for this experiment.}
\label{tab:added_windows}
\begin{tabular}{llrr}
\toprule
Task & Window & ACC & wF1 \\
\midrule
ADNI MCI/CN & TR0--39 & 62.39 & 62.69 \\
 & TR98--137 & 62.82 & 63.20 \\
ADNI AD/CN & TR0--39 & 75.46 & 75.41 \\
 & TR98--137 & $73.15\pm2.89$ & $73.35\pm3.10$ \\
\bottomrule
\end{tabular}
\end{table}

\subsection{Class Distribution and Probe Sensitivity}
\label{app:added_probes}

The HCP feature cohort used for these supplementary controls contains 1,000 subjects, with the class distribution shown in Table~\ref{tab:added_classes}. On this evaluation, the main MLP achieves $81.81\pm2.03$ balanced accuracy and $81.88\pm2.06$ macro F1, in addition to the reported accuracy and weighted F1.

\begin{table}[ht]
\centering\small
\caption{HCP class distribution for the supplementary frozen-feature evaluations. Class labels follow the saved split manifest.}
\label{tab:added_classes}
\begin{tabular}{lrrr}
\toprule
Split & Class 0 & Class 1 & Total \\
\midrule
Training & 321 & 281 & 602 \\
Validation & 107 & 94 & 201 \\
Test & 107 & 90 & 197 \\
\bottomrule
\end{tabular}
\end{table}

For Table~\ref{tab:added_probes}, we replace the MLP with a single trainable linear layer, retaining the frozen features, subject splits, training-set standardization, and validation-based checkpoint selection.

\begin{table}[ht]
\centering\small
\caption{\textbf{Probe sensitivity on resting-state tasks.} Accuracy / weighted F1 (\%), mean$\pm$std over three runs. PPMI uses the raw re-evaluation paired with the linear experiment.}
\label{tab:added_probes}
\resizebox{\linewidth}{!}{\begin{tabular}{lcc}
\toprule
Task & Main MLP & Linear probe \\
\midrule
HCP sex & $82.06\pm2.05$ / $82.04\pm2.05$ & $84.77\pm1.34$ / $84.77\pm1.34$ \\
ADNI MCI/CN & $61.11\pm1.48$ / $60.75\pm1.16$ & $50.43\pm6.45$ / $50.61\pm6.41$ \\
ADNI AD/CN & $75.46\pm2.12$ / $75.41\pm2.11$ & $70.37\pm10.79$ / $71.10\pm10.59$ \\
PPMI three-way & $54.51\pm5.35$ / $55.59\pm4.20$ & $44.10\pm0.60$ / $47.75\pm0.39$ \\
\bottomrule
\end{tabular}}
\end{table}

\subsection{Retrained and Fixed Readouts}
\label{app:added_readout}

For the retrained controls, a new MLP or linear probe is fitted independently for every input condition, using the same architecture, optimizer, splits, and seeds. For the fixed-readout controls, we train the classifier only on real-flatmap features and then apply it unchanged to perturbed features, retaining the real-training-set standardizer. The real-feature endpoint reproduces the original saved predictions before cross-condition evaluation.

Table~\ref{tab:added_readout} reports both readout protocols for the same three perturbations.

\begin{table}[ht]
\centering\small
\caption{\textbf{Readout controls on HCP.} Weighted F1 (\%), mean$\pm$std over seeds 42, 43, and 44. Retrained probes are fitted separately for each condition; fixed probes are fitted only on real flatmaps.}
\label{tab:added_readout}
\resizebox{\linewidth}{!}{\begin{tabular}{lrrrr}
\toprule
Input & Retrained MLP & Fixed MLP & Retrained linear & Fixed linear \\
\midrule
Real flatmap & $82.04\pm2.05$ & $82.04\pm2.05$ & $84.77\pm1.34$ & $84.77\pm1.34$ \\
Spatial block shuffle & $76.18\pm1.83$ & $28.65\pm0.00$ & $77.53\pm1.05$ & $32.18\pm3.48$ \\
Cortical-pixel permutation & $72.87\pm1.97$ & $28.65\pm0.00$ & $76.43\pm1.41$ & $28.65\pm0.00$ \\
Phase randomization & $64.49\pm2.14$ & $29.74\pm1.89$ & $65.87\pm3.09$ & $28.65\pm0.00$ \\
\bottomrule
\end{tabular}}
\end{table}

\subsection{Matched Cortical-Region Baselines on NSD}

The same-mask comparison in Table~\ref{tab:added_nsd_masks} uses the four NSD subjects and image splits from Section~\ref{sec:nsd_results}. ROI-linear averages responses within masked HCP-MMP parcels and fits a linear classifier; the small MLP uses adaptively pooled pixels from the masked flatmaps. FlatClip achieves higher weighted F1 than both baselines under both masks and higher mAP on the Visual mask. On NSDgeneral, ROI-linear and FlatClip obtain comparable mAP.

\begin{table}[t]
\centering\small
\caption{\textbf{Same-mask baselines on NSD.} Mean mAP and weighted F1 (\%) across the four evaluated subjects.}
\label{tab:added_nsd_masks}
\begin{tabular}{lrrrr}
\toprule
& \multicolumn{2}{c}{Visual} & \multicolumn{2}{c}{NSDgeneral} \\
\cmidrule(lr){2-3}\cmidrule(lr){4-5}
Method & mAP & wF1 & mAP & wF1 \\
\midrule
ROI-linear & 14.28 & 22.91 & \textbf{16.45} & 23.80 \\
Small MLP & 14.06 & 24.66 & 15.18 & 26.04 \\
FlatClip & \textbf{15.84} & \textbf{26.92} & 16.35 & \textbf{26.97} \\
\bottomrule
\end{tabular}
\end{table}

\subsection{Direct Acquired-Volume Renderings}
\label{app:added_volume}

\paragraph{HCP.}
The mean-projection control renders the nonzero-voxel axial mean of each $96\times96\times96$ volume as a $96\times96$ image. The slice-wise control retains each non-empty axial slice and encodes it independently. We mean-pool patch features within a slice, then average slice embeddings within each time point and the resulting embeddings across the first 40 TRs. Both produce a 768-dimensional subject feature for the same MLP protocol. The slice-wise method requires one image-encoder pass per slice rather than one per cortical frame; the retained HCP volumes contain a median of 77 non-empty slices per TR.

\begin{table}[ht]
\centering\small
\caption{\textbf{HCP direct-volume 40-frame results.} Accuracy and weighted F1 (\%), mean$\pm$std over three probe seeds.}
\label{tab:added_hcp_volume}
\begin{tabular}{lrr}
\toprule
Representation & ACC & wF1 \\
\midrule
Axial mean projection & $72.93\pm1.63$ & $72.81\pm1.75$ \\
FlatClip cortical flatmap & $82.06\pm2.05$ & $82.04\pm2.05$ \\
Independent slices, then temporal mean & $89.34\pm1.34$ & $89.32\pm1.34$ \\
\bottomrule
\end{tabular}
\end{table}

\paragraph{NSD subj01.}
We compare cortical flatmaps with a three-orthogonal-view mean-projection montage and independent axial slices. For the slice-wise condition, the three GLM-beta occurrences of each of the 10,000 images are first averaged voxel-wise. Each resulting volume has 80 non-empty axial slices. Each $96\times96$ slice is processed by frozen SigLIP2, and its patch features are pooled into one embedding; the 80 slice embeddings are then averaged into an image-level feature. The split contains 8,550 training, 450 validation, and 1,000 shared test images.

\begin{table}[ht]
\centering\small
\caption{\textbf{NSD subj01 direct-volume comparisons.} mAP and weighted F1 (\%). Original FlatClip and mean-projection values are reported alongside the three-seed slice-wise experiment.}
\label{tab:added_nsd_volume}
\begin{tabular}{lrr}
\toprule
Input & mAP & wF1 \\
\midrule
FlatClip, whole cortex & 11.100 & 24.576 \\
FlatClip, Visual & 14.050 & 25.829 \\
FlatClip, NSDgeneral & 15.311 & 26.922 \\
Three-view mean projection & 4.95 & 17.00 \\
Independent axial slices & $4.342\pm0.171$ & $15.922\pm0.579$ \\
\bottomrule
\end{tabular}
\end{table}

\subsection{Omni-fMRI with Native Temporal Inputs}
\label{app:added_omni}

We evaluate the released Omni-fMRI checkpoint using official NSD subj01 1.8-mm preprocessed BOLD time series. Each target image has three presentations. We construct two 40-frame inputs:
\begin{enumerate}
\item \textbf{Continuous BOLD40.} A complete continuous 40-TR window is encoded for each presentation. The three presentation-level CLS features are averaged to obtain one feature per image.
\item \textbf{Event stack.} Response-centered 13-TR segments from the three presentations are concatenated and padded with one zero channel. The complete $3\times13+1$ stack is encoded once to produce an image-level CLS feature.
\end{enumerate}

Both temporal constructions use 10,000 images with a 7,000/1,000/2,000 training/validation/test split. We train a class-weighted BCE MLP with seeds 42, 43, and 44, select checkpoints by validation weighted F1, and evaluate the test set with a fixed threshold of 0.5. NSD's rapid event-related design produces overlapping BOLD responses: continuous windows retain acquisition order, whereas event stacks concatenate segments from separate presentations.

Table~\ref{tab:added_omni} reports the two temporal experiments alongside the original beta-map results. The original FlatClip row uses the 8,550/450/1,000 image split and validation-mAP selection; the temporal rows use the shared protocol specified above.

\begin{table}[ht]
\centering\small
\caption{\textbf{Omni-fMRI temporal input adaptations on NSD subj01.} mAP and weighted F1 (\%). Original beta-map results and new temporal-input results are grouped separately. Temporal rows report mean$\pm$std over three seeds.}
\label{tab:added_omni}
\begin{tabular}{lrr}
\toprule
Input & mAP & wF1 \\
\midrule
\multicolumn{3}{l}{\textit{Original beta-map results}} \\
FlatClip--NSDgeneral & 15.311 & 26.922 \\
Omni-fMRI, repeated beta $\times40$ & 3.949 & 13.818 \\
\midrule
\multicolumn{3}{l}{\textit{Native temporal inputs: 7,000/1,000/2,000 split}} \\
Omni-fMRI, continuous BOLD40 & $3.804\pm0.052$ & $16.531\pm0.357$ \\
Omni-fMRI, event stack $3\times13+1$ & $3.840\pm0.035$ & $16.188\pm0.183$ \\
\bottomrule
\end{tabular}
\end{table}

\subsection{Labeled-Subject Scaling}
\label{app:added_scaling}

We use cached HCP features and keep the validation and test subjects in Table~\ref{tab:added_classes} fixed. Class-stratified subsets contain 10\%, 25\%, 50\%, or 100\% of the training subjects and are nested within each seed. Standardization is fitted on each training subset, and checkpoints are selected by validation weighted F1. Table~\ref{tab:added_scaling} reports both probes at each training fraction.

\begin{table}[ht]
\centering\small
\caption{\textbf{HCP labeled-subject scaling.} Accuracy / weighted F1 (\%), mean$\pm$std over three seeds.}
\label{tab:added_scaling}
\resizebox{\linewidth}{!}{\begin{tabular}{lrcc}
\toprule
Training fraction & Subjects & Linear & MLP \\
\midrule
10\% & 60 & $70.05\pm4.51$ / $70.01\pm4.47$ & $70.39\pm2.55$ / $69.92\pm3.04$ \\
25\% & 150 & $77.83\pm2.55$ / $77.83\pm2.59$ & $75.97\pm4.93$ / $75.89\pm4.96$ \\
50\% & 300 & $82.91\pm0.29$ / $82.89\pm0.27$ & $79.19\pm1.76$ / $79.11\pm1.71$ \\
100\% & 602 & $84.77\pm1.34$ / $84.77\pm1.34$ & $82.06\pm2.05$ / $82.04\pm2.05$ \\
\bottomrule
\end{tabular}}
\end{table}

\end{document}